\documentclass[aps, twocolumn,floats, showpacs]{revtex4}
\usepackage{amsmath,amssymb,graphicx}
\usepackage{epsfig}
\usepackage{graphicx}
\usepackage{enumerate}

\newcommand{\avg}[1]{\left< #1 \right>} % for average
\renewcommand{\d}[2]{\frac{d #1}{d #2}} % for derivatives
\newcommand{\dd}[2]{\frac{d^2 #1}{d #2^2}} % for double derivatives
\newcommand{\pd}[2]{\frac{\partial #1}{\partial #2}} % for partial derivatives
\newcommand{\ket}[1]{\left| #1 \right>} % for Dirac bras
\newcommand{\bra}[1]{\left< #1 \right|} % for Dirac kets
\newcommand{\intinf}{\int_{-\infty}^\infty}
\newcommand{\beq}{\begin{equation}}
\newcommand{\eeq}{\end{equation}}

\newcommand{\bea}{\begin{eqnarray}}
\newcommand{\eea}{\end{eqnarray}}

\begin{document}

%\preprint{version 7}

%\baselineskip 10mm

\title{Vibrational cooling, heating, and instability in
molecular conducting junctions: Full counting statistics analysis}
\author{Lena Simine}
\affiliation{Chemical Physics Theory Group, Department of Chemistry, University of Toronto,
80 Saint George St. Toronto, Ontario, Canada M5S 3H6}
\author{Dvira Segal}
\affiliation{Chemical Physics Theory Group, Department of Chemistry,
University of Toronto, 80 Saint George St. Toronto, Ontario, Canada
M5S 3H6}

\date{\today}
\begin{abstract}
We study current-induced vibrational cooling, heating, and
instability in a donor-acceptor rectifying molecular junction using
a full counting statistics approach. In our model, electron-hole
pair excitations are coupled to a given molecular vibrational mode
which is either harmonic or highly anharmonic. This mode may be
further coupled to a dissipative thermal environment. Adopting a master
equation approach, we confirm the charge and heat exchange
fluctuation theorem in the steady-state limit, for both harmonic and
anharmonic models. Using simple analytical expressions, we calculate
the charge current
%its noise,
and several measures for the mode effective temperature. At low
bias, we observe the effect of bias-induced cooling of the
vibrational mode. At higher bias, the mode effective temperature is
higher than the environmental temperature, yet the junction is
stable. Beyond that, once the vibrational mode (bias-induced)
excitation rate overcomes its relaxation rate, instability occurs.
We identify regimes of instability as a function of voltage bias and
coupling to an additional phononic thermal bath. Interestingly, we
observe a reentrant behavior where an unstable junction can properly
behave at a high enough bias. The mechanism for this behavior is
discussed.
\end{abstract}

%\pacs{05.70.-a, 05.70.Ln, 89.70.-a, 89.70.Kn}

\maketitle

%-------------------

% Introduction
\section{Introduction}

% Heating in mol elec
Can molecules serve as reliable components in electronic circuits? A
major obstacle in realizing molecular-based electronic devices is
junction heating and breakdown, the result of vibrational excitation
by the electron current \cite{ExpVib1, ExpVib2, McEuen,
Selzer,Heating3, HeatSegal, Heating1,Heating2,NitzanVib,Thoss}. This
situation generally occurs once the bias voltage exceeds typical
molecular vibrational frequencies and the electronic levels are
situated within the bias window. If energy dissipation from the
conducting object to its environment (metals, solvent) is not
efficient, the molecular conductor experiences significant heating,
ultimately leading to junction breakdown. A related question, the
possibility for a nonequilibrium induced cooling of the junction has
been the topic of recent experimental and theoretical studies
\cite{Cooling,Selzer,CoolingG,ThossC}.

\begin{figure}[htbp]
\vspace{-3mm}  {\hbox{\epsfxsize=120mm
\hspace{10mm}\epsffile{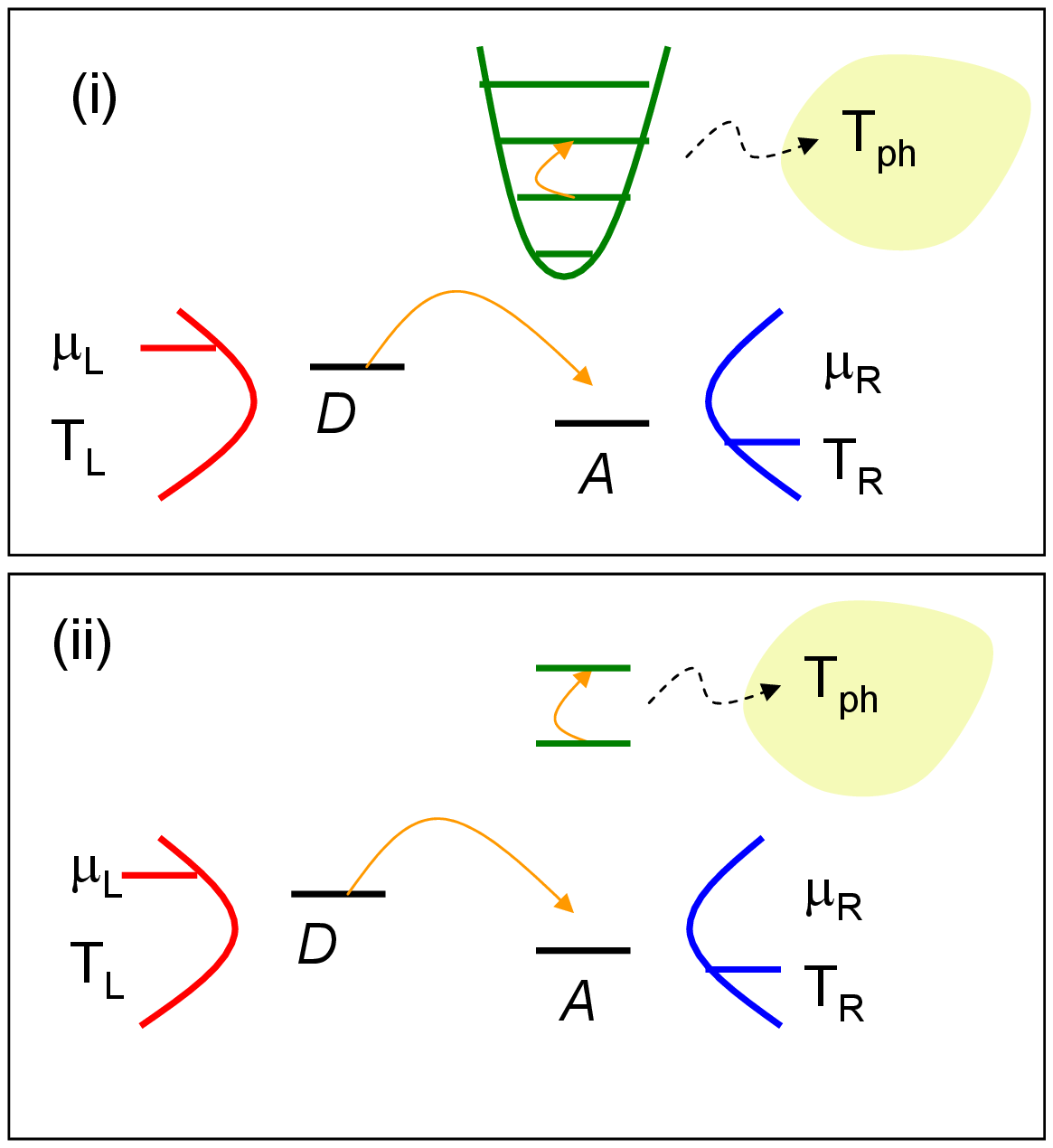}}} \vspace{-25mm}\caption{
Schemes of the two models considered in this work.
A biased donor-acceptor electronic junction is
coupled to either (i) a harmonic molecular mode, or (ii) an anharmonic 
mode, represented by a two-state system.
In both cases the molecular mode may further relax its energy
to a phononic thermal reservoir,
maintained at the temperature $T_{ph}$.
This coupling is represented by a dashed arrow.
}
\label{FigS}
\end{figure}

% Setup here
In this paper, we study the problem of bias-induced molecular
cooling, heating, and (potential) junction breakdown due to
vibrational instabilities, using the Donor (D)-Acceptor (A)
Aviram-Ratner electronic rectifier setup \cite{Aviram}, see Fig.
\ref{FigS}. By coupling electronic transitions within the junction
to a particular internal molecular vibrational mode, significant
molecular heating can take place once the donor level is lifted
above the acceptor level, as the excess electronic energy is
used to excite the vibrational mode. This process may ultimately
lead to junction instabilities and breakdown \cite{Lu}. The
model can also demonstrate current-induced cooling at low bias, when
tuning the junction's parameters.

% mol-el issue of interest
Within this simple system, several issues are of fundamental and
practical interest. First, one would like to understand the role of
mode anharmonicity in the transport process and in the heating or
cooling behavior. Second, the molecular vibration under
investigation, the one controlling junction stability, can be
assumed to be  well isolated from other modes. Alternatively, this
mode may be coupled to other phonons, allowing for energy damping to
a larger environment. These two situations should result in
distinctive cooling or heating behaviors. These issues will be
explored here. Other relevant challenges which are not considered
here are the possibility to selectively excite vibrational modes in
the molecule, using voltage bias \cite{Peskin}, or more generally, to
drive molecular motion or trigger chemical dynamics \cite{Seideman}.

%================================
% scheme
%================================

% FT
Using a full counting statistics (FCS) approach, our analysis
further  contributes to the institution of fluctuation relations in
open many-body quantum systems. Fluctuation theorems (FT) for
entropy production quantifies the probability of negative entropy
generation, measuring ``second law violation" \cite{Evans,Cohen}.
Such ``anomalous" processes are relevant at the nanoscale. While
originally demonstrated in classical systems \cite{EFluctC}, recent
experimental efforts are dedicated to explore their validity within
quantum systems \cite{EFluctQ}.
From the theoretical side, the extension of the work and heat FT to
the quantum domain has recently attracted significant attention
\cite{Mukamel-rev, Hanggi-rev}. Specifically, a quantum exchange FT,
for the transfer of charge and energy between two reservoirs
maintained at different chemical potentials and temperatures, has
been derived in Ref. \cite{Tasaki} using projective measurements,
and in Ref. \cite{Mukamel-rev} based on the unraveling of the
quantum master equation. It is of interest to test these relations
in particular cases, e.g., for systems strongly coupled to multiple
reservoirs, when the reservoirs cooperatively affect the subsystem
\cite{FT1}, including nonmarkovian reservoirs \cite{FT2,Emary,Braggio}, and
in models showing coupled charge and energy transfer processes, yet
the respective fluxes are not tightly coupled.
%and the two reservoirs cooperatively damp the subsystem energy.
The system investigated here corresponds to the latter case.

% Method and Goals
Different flavors of the phonon-assisted-tunneling model have been
analyzed in the literature \cite{NitzanVib}. Among the various
techniques adopted we list solution of the dynamics as a
scattering problem \cite{Bonca}, extension of the basic
nonequilibrium Green's function formalism to include molecular
vibrations \cite{Galperin}, or the use of master equation approaches
\cite{Mitra}.
In this paper, we exploit the latter method, and present a
full-counting statistics of the system, allowing for the exploration
of charge current, energy current and noise processes at the same
footing. Further, we {\it analytically} obtain the cumulant
generating function (CGF) of the model, allowing for the
verification of the steady-state charge-energy fluctuation theorem
in this many-body quantum system.

The objectives of this work are therefore twofold: (i) to analyze a
simple model that can elucidate cooling, heating and instability
mechanisms in molecular rectifiers, specifically, to understand the
roles of mode anharmonicity and additional damping routes, and (ii) to
establish the steady-state entropy production fluctuation theorem
within a nonequilibrium quantum model, transferring charge and
energy between the reservoirs in a cooperative manner. Recent
studies have analyzed the role of electron-vibration interaction on
the full counting statistics (FCS) within different approaches
%the quantum master equation approach
\cite{Novotny,Asai,Emary,GalperinFCS,Komnik}.
Complementing these efforts, our treatment offers an analytic structure for the
CGF, allowing for a clear inspection of the microscopic processes
involved.

The plan of the paper is as follows. In Sec. II we introduce the D-A
molecular rectifier and its two flavors, either including a harmonic
or an anharmonic internal vibration. In Sec. III the anharmonic
model is analyzed within a FCS approach, demonstrating cooling,
heating and instability dynamics at different parameter regions. The case
with an additional phonon bath is considered in Appendix A. Sec. IV
explores the harmonic mode model. Sec. V concludes.

%================================

\section{Model}

Our model includes a biased molecular electronic junction and a
selected internal vibrational mode which is coupled to an electronic
transition in the junction. This mode possibly interacts with other
(reservoir) phonons, an extension presented in Appendix A. For a
schematic representation, see Fig. \ref{FigS}.
%Generally, the model describes an electron-hole pair excitation by molecular
%vibrations.
Generally, this model allows one to investigate the exchange of
electronic energy with molecular (vibrational) heating. The total
Hamiltonian is given by the following terms,
\bea H= H_{M}+H_L +H_R+ H_{c}+ H_{vib} + H_I. \label{eq:H} \eea
The first term, $H_M$, stands for the molecular electronic part
including two electronic states
\bea
H_M=\epsilon_dc_d^{\dagger}c_d + \epsilon_ac_a^{\dagger}c_a.
\eea
Here, $c_{d/a}^{\dagger}$ ($c_{d/a}$) is a fermionic creation
(annihilation) operator of an electron on the donor or acceptor
sites, of energies $\epsilon_{d,a}$. The second and third terms
in Eq. (\ref{eq:H}) describe the two metals, $H_{\nu}$, $\nu=L,R$,
each including a collection of noninteracting electrons
\bea H_{L}=\sum_{l\in L} \epsilon_{l}
c_l^{\dagger}c_l;\,\,\,\,\,\,\,H_{R}=\sum_{r\in R} \epsilon_{r}
c_r^{\dagger}c_r. \eea
The hybridization of the donor state to the left
($L$) bath, and similarly, the coupling of the acceptor site to the
right ($R$) metal, are incorporated into $H_c$,
\bea H_c= \sum_{l} v_l\left(c_l^{\dagger}c_d +
c_d^{\dagger}c_l\right) + \sum_{r} v_r\left( c_r^{\dagger}c_a +
c_a^{\dagger}c_r\right).
 \eea
%
%The coupling energy $v_{l/r}$ is assumed to be strong. XXX
The Hamiltonian further includes an  internal molecular vibrational mode
of frequency $\omega_0$. The mode displacement from equilibrium is
coupled to an electron hopping in the system with an energy cost
$\kappa$, resulting in heating and/or cooling effects,
\bea H_{vib}&=& \omega_0 b_0^{\dagger}b_0,
\nonumber\\
H_{I}&=& \kappa \left[c_d^{\dagger}c_a + c_a^{\dagger}c_d
\right](b_0^{\dagger}+b_0). \eea
Here, $b_0^{\dagger}$ ($b_0$) represents a bosonic creation
(annihilation) operator. Note that in our construction the donor and
acceptor sites are coupled to each other only through the
interaction with the vibrational mode. We now diagonalize the
electronic part of the Hamiltonian, $H_{el}=H_M+H_L+H_R+H_c$, to
obtain, separately, the exact eigenstates for the $L$-half and
$R$-half of $H_{el}$,
\bea H_{el}
%&=& \tilde H_{L}+\tilde H_R \nonumber\\
%\tilde H_L&=&
=\sum_{l} \epsilon_l a_{l}^{\dagger}a_l+
%\tilde H_R=
\sum_{r}\epsilon_r a_{r}^{\dagger}a_r. \eea
Assuming that the reservoirs are dense, their new operators
are assigned energies same as those before diagonalization.
The donor and acceptor (new) energies are assumed to be placed
within the band of continuous states, excluding the existence of
bound states. The old operators are related to the exact eigenstates
by \cite{Mahan}
\bea c_d&=&\sum_{l}\lambda_l a_l, \,\,\,\,\,\,\
c_l=\sum_{l'}\eta_{l,l'}a_{l'} \nonumber\\
c_a&=&\sum_{r}\lambda_r a_r, \,\,\,\,\,\,\
c_r=\sum_{r'}\eta_{r,r'}a_{r'},
 \eea
where the coefficients, e.g., for the $L$ set, are given by
\bea \lambda_l&=&\frac{v_l}{\epsilon_l-\epsilon_d-\sum_{l'}
\frac{v_{l'}^2}{\epsilon_l-\epsilon_{l'}+i\delta}} \nonumber\\
\eta_{l,l'}&=&\delta_{l,l'}-\frac{v_l\lambda_{l'}}{\epsilon_l-\epsilon_{l'}+i\delta}.
\label{eq:diag}
 \eea
Similar expressions hold for the $R$ set. It is easy to derive
the following relation,
\bea
\sum_{l'}\frac{v_{l'}^2}{\epsilon_{l}-\epsilon_{l'}+i\delta}=
PP \sum_{l'}\frac{v_{l'}^2}{\epsilon_l-\epsilon_{l'}} -i \Gamma_L(\epsilon_l)/2,
\eea
with the hybridization strength %$\Gamma=\Gamma_L+\Gamma_R$,
$\Gamma_L(\epsilon)=2\pi \sum_l v_l^2\delta(\epsilon-\epsilon_l)$.
The expectation values of the exact eigenstates are
\bea \langle a_l^{\dagger}a_{l'}
\rangle=\delta_{l,l'}f_L(\epsilon_l),\,\,\,\,\, \langle
a_r^{\dagger}a_{r'} \rangle=\delta_{r,r'}f_R(\epsilon_r),
\eea
where $f_L(\epsilon)=[\exp(\beta_L(\epsilon-\mu_L))+1]^{-1}$ denotes
the Fermi distribution function. An analogous  expression holds for
$f_R(\epsilon)$. The reservoirs temperatures are denoted by
$1/\beta_{\nu}$ ; the chemical potentials are  $\mu_{\nu}$. With the
new operators, the Hamiltonian (\ref{eq:H}) can be rewritten as
\bea H_{H} &=& \sum_l  \epsilon_l a_l^{\dagger}a_l + \sum_r \epsilon_r
a_r^{\dagger}a_r + \omega_0 b_0^{\dagger}b_0 \nonumber\\ &+& \kappa
\sum_{l,r} \left[\lambda_l^* \lambda_r a_l^{\dagger}a_r +
\lambda_r^* \lambda_l a_r^{\dagger}a_l\right] (b_0^{\dagger}+b_0).
\label{eq:HTA} \eea
In this form, the model generally describes the process of an electron-hole pair
excitation by a molecular vibration.
We denote it by $H_H$, to highlight the vibrational mode harmonicity.
A simple version of the model is reached by
replacing the harmonic mode by a two-state system (spin),
using the Pauli matrices,
\bea  H_{A} &=& \sum_l  \epsilon_l a_l^{\dagger}a_l + \sum_r
 \epsilon_r a_r^{\dagger}a_r + \frac{\omega_0}{2} \sigma_z \nonumber\\
&+& \kappa \sum_{l,r} \left[\lambda_l^* \lambda_r a_l^{\dagger} a_r
+ \lambda_r^* \lambda_l a_r^{\dagger}a_l\right] \sigma_x.
\label{eq:HTB} \eea
%
%This simplified model is referred to as the ``two-level system
%electron-hole excitation (TLSEH) model``.
The truncated harmonic spectrum imitates an anharmonic mode, as
only several (two in the present extreme case) states are
bounded within the anharmonic potential \cite{SegalM}. We denote
this Hamiltonian by $H_A$, to indicate on the anharmonicity of the
molecular mode. The dynamics of this model should coincide with the
behavior dictated by $H_H$, at low temperatures.

Charge and energy transfer dynamics in these models can be followed
by studying electronic properties  \cite{Gernot,GalperinFCS,Komnik}.
In contrast, here we explore the junction response to an applied
voltage bias by studying the vibrational mode excitation and
relaxation dynamics. The analysis of the two-state model, Eq.
(\ref{eq:HTB}), therefore turns out to be {\it simpler} than the
case when the vibrational mode has an infinite spectrum. In what
follows, we derive in details the CGF for the anharmonic-mode case.
Appendix A generalizes this calculation to include an additional
dissipative thermal bath. We then extend these results and discuss
the model conveyed by Eq. (\ref{eq:HTA}).

%=============================
% rates
\begin{figure}[htbp]
\vspace{0mm} \hspace{0mm} {\hbox{ \hspace{10mm}\epsfxsize=120mm
\epsffile{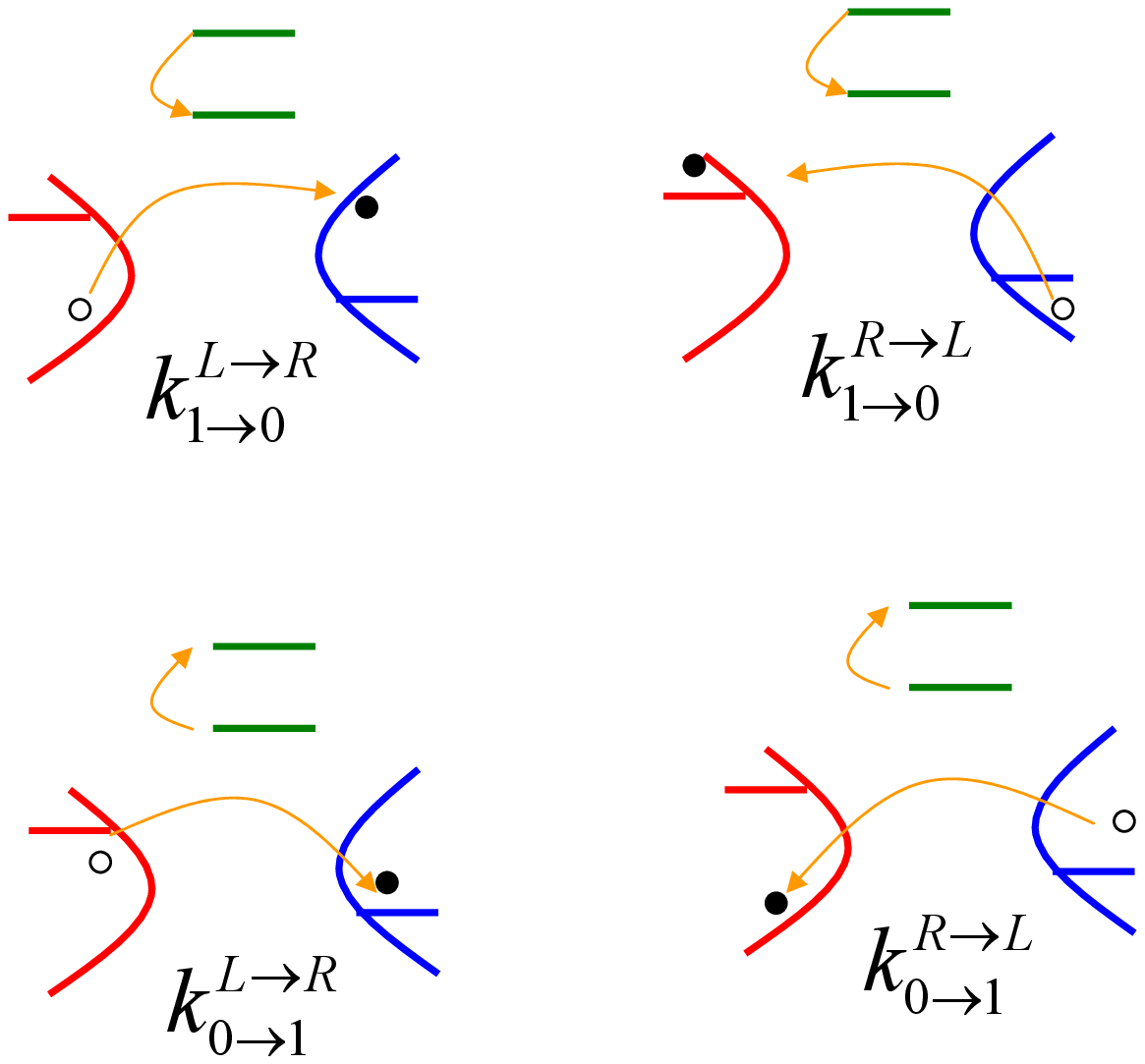}}} \vspace{-40mm}\caption{Scheme of the vibrational mode 
excitation and relaxation processes.
%taking place within the anharmonic mode.
A full circle represents an electron transferred; a hollow circle depicts
the hole that has been left behind.
} \label{FigR}
\end{figure}
%============================

\section{Anharmonic-mode Rectifier}

\subsection{Impurity dynamics}

We explore the dynamics of an anharmonic mode,
referred to as an ``impurity", or a two-state-system (TLS),
within an  electronic rectifier,
assuming a weak donor-acceptor - mode interaction.
We rewrite Eq. (\ref{eq:HTB}) as
\bea H_{A} = \frac{\omega_0}{2} \sigma_z + \sigma_x F_e + \sum_l
\epsilon_l a_l^{\dagger}a_l + \sum_r  \epsilon_r a_r^{\dagger}a_r,
\label{eq:HBB} \eea
by defining the electron-hole pair generation operator as
\bea F_e=\kappa \sum_{l,r} (\lambda_l^* \lambda_r a_l^{\dagger}a_r +
\lambda_r^* \lambda_l a_r^{\dagger}a_l).
\label{eq:Fe}
\eea
The Hamiltonian (\ref{eq:HBB}) can be transformed to the spin-fermion model
\cite{MitraSpin, SMarcus} of zero energy spacing,
using the unitary transformation
\bea
U^{\dagger}\sigma_z U=\sigma_x, \,\,\,\,\,  U^{\dagger}\sigma_x U=\sigma_z,
\eea
with $U=\frac{1}{\sqrt{2}}(\sigma_x+\sigma_z)$.
The transformed Hamiltonian $\tilde H_A=U^{\dagger}H_AU$ is given by
\bea \tilde H_{A} = \frac{\omega_0}{2} \sigma_x + \sigma_z F_e + \sum_l
\epsilon_l a_l^{\dagger}a_l + \sum_r  \epsilon_r a_r^{\dagger}a_r.
 \eea
In this form, the TLS dynamics can be simulated {\it exactly} using an influence-functional
 path integral approach \cite{IF}.

Back to (\ref{eq:HBB}),
we denote the TLS ground state and excited state by  $|0\rangle$ and
 $|1\rangle$, with energies $0$ and $\omega_0$,
respectively. We express the Pauli operators by these states,
$\sigma_z=|1\rangle \langle 1| - |0\rangle \langle 0|$,
$\sigma_x=|1\rangle \langle 0| + |0\rangle \langle 1|$.
Next, using the quantum Liouville equation,
we obtain kinetic-rate equations for the states population
$p_{n}$ ($n$=0,1)  \cite{Breuer,SegalM}. This standard derivation involves a
 second order perturbation theory treatment with respect to $\kappa$,
the mode-molecule coupling parameter, followed by a Markov
approximation. The resulting equation for the reduce density matrix $\rho_S$ take the simple form
($\hbar\equiv 1$)
\bea
\dot \rho_S&=& -i[V(t),\rho_S(0)] 
\nonumber\\
&-& 
\int_0^{\infty} d\tau {\rm Tr}_{B}\left\{  [ V(t),[V(\tau),\rho_S(t) \rho_L\rho_R]  ]\right\}
\eea
Here, $V=\sigma_x F_e$ represents the (mode-molecule) coupling term in Eq. (\ref{eq:HBB}).
The operators are written in the interaction representation, 
$O(t)=e^{i (H_A-V) t}Oe^{-i (H_A-V)t}$ and we trace over the electronic degrees of freedom.
The reservoirs $\nu=L,R$ are maintained in a grand canonical state as
$\rho_{\nu}=e^{-\beta_{\nu} (H_{\nu}-\mu_{\nu}N_{\nu})}/Z_{\nu}$;
$Z_{\nu}$ is the partition function of the $\nu$ bath. 
Identifying the diagonal matrix elements as population, $p_n=[\rho_S]_{n,n}$, we obtain the kinetic equation
\bea
\dot p_1=-k_{1\rightarrow 0}^e p_1 +k_{0\rightarrow 1}^ep_0,\,\,\,\,\ p_1+p_0=1.
\label{eq:DTLS}
\eea
In this model, the off-diagonal elements of the reduced density
matrix are naturally decoupled from the population dynamics
\cite{Silbey}. The excitation ($k_{0\rightarrow 1}^e$) and relaxation
($k_{1\rightarrow 0}^e$) rate constants are given by Fourier
transforms of bath correlation functions 
\bea k_{1\rightarrow0}^e &=& \int_{-\infty}^{\infty} e^{i\omega_0 \tau} \langle F_e(\tau) F_e(0)\rangle d\tau
\nonumber\\
k_{0\rightarrow1}^e &=& \int_{-\infty}^{\infty} e^{-i\omega_0 \tau}
\langle F_e(\tau) F_e(0)\rangle d\tau, \label{eq:kTLS} \eea
enclosing electron-hole pair excitation processes,
\bea
\langle F_e(t)F_e(0)\rangle &=& \kappa^2
{\rm Tr}_L{\rm Tr}_R \Big\{
\sum_{l,l'}\sum_{r,r'} \rho_L \rho_R
\nonumber\\
&\times&
\Big[\lambda_l^*\lambda_r a_l^{\dagger}(t)a_r(t)
+\lambda_r^*\lambda_l a_r^{\dagger}(t)a_l(t)\Big]
\nonumber\\
&\times& \Big[\lambda_{l'}^*\lambda_{r'}a
_{l'}^{\dagger}(0)a_{r'}(0) +\lambda_{r'}^*\lambda_{l'}
a_{r'}^{\dagger}(0)a_{l'}(0)\Big] \Big\}.
 \nonumber\\ \eea
The operators are given in the interaction representation, e.g.,
$a_l^{\dagger}(t)=e^{i H_L t}a_l^{\dagger}e^{-i H_Lt}$. 
As we separately trace over the $L$ and $R$-baths' degrees of freedom,
it can be shown that the rate constants can be decomposed into two
contributions,
\bea
k_{1\rightarrow 0}^e=k_{1\rightarrow 0}^{L\rightarrow
R}+k_{1\rightarrow 0}^{R\rightarrow L};\,\,\,\ k_{0\rightarrow
1}^e=k_{0\rightarrow 1}^{L\rightarrow R}+k_{0\rightarrow
1}^{R\rightarrow L},  \eea
satisfying
\bea
 k_{1\rightarrow0}^{L\rightarrow R}&=&
2\pi\kappa^2 \nonumber\\ &\times &
\sum_{l,r}|\lambda_l|^2|\lambda_r|^2f_L(\epsilon_l)(1-f_R(\epsilon_r))\delta(\omega_0+\epsilon_l-\epsilon_r)
%k_{1\rightarrow0}^{R\rightarrow L}&=&2\pi
% \kappa^2\sum_{l,r}|\lambda_l|^2|\lambda_r|^2f_R(\epsilon_r)(1-f_L(\epsilon_l))\delta(\omega_0+\epsilon_r-\epsilon_l)
\nonumber\\
 k_{0\rightarrow 1}^{L\rightarrow R}&=&
2\pi \kappa^2
\nonumber\\ &\times& \sum_{l,r}|\lambda_l|^2|\lambda_r|^2f_L(\epsilon_l)(1-f_R(\epsilon_r))\delta(-\omega_0+\epsilon_l-\epsilon_r).
%\nonumber\\
%k_{0\rightarrow 1}^{R\rightarrow L}&=& 2\pi\kappa^2
%\sum_{l,r}|\lambda_l|^2|\lambda_r|^2f_R(\epsilon_r)(1-f_L(\epsilon_l))\delta(-\omega_0+\epsilon_r-\epsilon_l)
\nonumber\\
\label{eq:rate1}
 \eea
Similar relations hold for the right-to-left going excitations. 
The energy in the Fermi function $f_{\nu}(\epsilon)$  is measured with respect to
the (equilibrium) Fermi energy, placed at $(\mu_L+\mu_R)$, and we assume that the bias
is applied symmetrically, $\mu_L=-\mu_R$. The four rate constants describe distinct
electron-hole excitation processes, depicted in
Fig. \ref{FigR}. At forward bias, if we set the effective density of states (DOS) of the $L$ bath to lie
higher in energy that the DOS of the right bath, we immediately note that
the rate $k_{0\rightarrow 1}^{L\rightarrow R}$
should dominate over $k_{1\rightarrow 0}^{L\rightarrow R}$, potentially leading to "population inversion"
of the vibrational mode.
Utilizing electronic reservoirs with energy dependent DOS 
is thus the basic ingredient of the instability formation here, as we show below.
For convenience, we define the spectral density for
the $\nu$ bath as
\bea J_{\nu}(\epsilon)&=&2\pi
\kappa\sum_{j\in\nu}|\lambda_{j}|^2\delta(\epsilon_{j}-\epsilon).
\eea
Explicitly, using Eq. (\ref{eq:diag}), we find that
this function has a Lorentzian lineshape, and that it is centered around either the D or A level,
\bea
J_L(\epsilon)&=&\kappa\frac{\Gamma_L(\epsilon)}{(\epsilon-\epsilon_d)^2+\Gamma_L(\epsilon)^2/4}
\nonumber\\ J_R(\epsilon)&=&
\kappa\frac{\Gamma_R(\epsilon)}{(\epsilon-\epsilon_a)^2+\Gamma_R(\epsilon)^2/4}.
\label{eq:spec}
 \eea
Using the spectral density function,
we express the terms in Eq. (\ref{eq:rate1}) as integrals
\bea
 k_{1\rightarrow0}^{L\rightarrow R}&=&\frac{1}{2\pi}
\int_{-\infty}^{\infty}
f_L(\epsilon)\left[1-f_R(\epsilon+\omega_0)\right]
J_L(\epsilon)J_R(\epsilon+\omega_0)d\epsilon
\nonumber\\
k_{1\rightarrow0}^{R\rightarrow L}&=&\frac{1}{2\pi}\int_{-\infty}^{\infty}
 f_R(\epsilon)\left[1-f_L(\epsilon+\omega_0)\right]
J_R(\epsilon)J_L(\omega_0+\epsilon)d\epsilon
\nonumber\\
 k_{0\rightarrow 1}^{L\rightarrow R}&=&\frac{1}{2\pi}
\int_{-\infty}^{\infty}
f_L(\epsilon)\left[1-f_R(\epsilon-\omega_0)\right]
J_L(\epsilon)J_R(\epsilon-\omega_0)d\epsilon
\nonumber\\
k_{0\rightarrow 1}^{R\rightarrow L}&=&\frac{1}{2\pi}
\int_{-\infty}^{\infty}f_R(\epsilon)\left[1-f_L(\epsilon-\omega_0)\right]
J_R(\epsilon)J_L(\epsilon-\omega_0)d\epsilon.
\nonumber\\
\label{eq:rate2}
 \eea
The following relations hold ($\beta_L=\beta_R$
and $\Delta \mu\equiv \mu_L-\mu_R$),
\bea
\frac{ k_{1\rightarrow 0}^{R\rightarrow L}} {k_{0\rightarrow 1}^{L\rightarrow R}}
=e^{-\beta \Delta \mu} e^{\beta\omega_0}
; \,\,\,\,\,
\frac{k_{1\rightarrow 0}^{L\rightarrow R}} {k_{0\rightarrow 1}^{R\rightarrow L}}
=e^{\beta \Delta \mu} e^{\beta\omega_0}.
\eea
In equilibrium, detailed bias is therefore maintained.

The dynamics conveyed by Eqs. (\ref{eq:DTLS})-(\ref{eq:rate2}) is
non-separable in terms of the two metals, in contrast to simple
linear interaction cases \cite{SegalM}. In other words, the reservoirs
cooperatively excite or damp energy from the impurity,
thus their action is non-additive.

It should be noted that while we assume a weak interaction limit, between
electron-hole pair generation and the vibrational mode, our scheme
does {\it not} enforce weak metal-molecule coupling; this part is
exactly diagonalized to yield the reservoirs spectral function,
peaked about the D or A levels. If one where to force weak
metal-molecule interaction, the spectral functions (\ref{eq:spec})
would reduce to delta functions, $J_L(\epsilon)=2 \pi \kappa
\delta(\epsilon-\epsilon_d)$ and $J_R(\epsilon)=2 \pi \kappa
\delta(\epsilon-\epsilon_a)$, and the resulting rates would be
evaluated at the donor and acceptor levels, e.g.,
$k_{1\rightarrow 0}^{L\rightarrow R}=2\pi\kappa^2f_L(\epsilon_d)[1-f_R(\epsilon_a)]\delta(\epsilon_d-\epsilon_a+\omega_0).
$
This also implies that charge and energy currents are not ``tightly coupled" here,
such that for each transferred electron not necessarily precisely one quanta of energy should
be gained or drained at either contact.
In this aspect, our study complements the work reported in \cite{Thoss}. There, using the
small polaron transformation, the coupling of the molecular bridge
to the leads is assumed to be weak, while its coupling to the vibrational mode can be made large.
%While the FT for charge and energy transfer has been generally tested
%for linear interaction models \cite{Mukamel-rev}, in what follows we
%confirm its validity for the present, more complex situation.

%========================================================
\subsection{Resolved charge and energy equations}

We write here a closed expression for the cumulant generating function,
following the approach developed in Refs. \cite{FT1,FT2}. It will
allow us to obtain the current, its noise power, and to confirm the FTs in this system.
We define $\mathcal P_t(n,N,\omega)$ as the probability that by the
time $t$ the impurity (TLS) occupies the state $n$, $N$ electrons
have been transferred from the $L$ metal to the $R$ side, and a net
energy $\omega$ has been transferred, $L$ to $R$. Resolving Eq.
(\ref{eq:DTLS}) to its charge and energy components, we find that
this probability satisfies the following equation of motion
\cite{FT1,FT2},
\begin{widetext}
\bea
 \dot{\mathcal  P}_t(1,N,\omega)&=&
- \mathcal P_t(1,N,\omega) k_{1\rightarrow 0}^e
\nonumber\\
&+& \intinf \mathcal P_t(0,N-1, \omega-\epsilon +\omega_0)
f_L(\epsilon) [1-f_R(\epsilon-\omega_0)]
J_L(\epsilon)J_R(\epsilon-\omega_0) d\epsilon
\nonumber\\
&+&\intinf \mathcal P_t(0,N+1, \omega+\epsilon) f_R(\epsilon)
[1-f_L(\epsilon-\omega_0)] J_R(\epsilon)J_L(\epsilon-\omega_0)
d\epsilon
\nonumber\\
 \dot{\mathcal  P}_t(0,N,\omega)&=&
- \mathcal P_t(0,N,\omega) k_{0\rightarrow 1}^e
\nonumber\\
&+& \intinf \mathcal P_t(1,N-1, \omega-\epsilon -\omega_0)
f_L(\epsilon) [1-f_R(\epsilon+\omega_0)]
J_L(\epsilon)J_R(\epsilon+\omega_0) d\epsilon
\nonumber\\
&+&\intinf \mathcal P_t(1,N+1, \omega+\epsilon) f_R(\epsilon)
[1-f_L(\epsilon+\omega_0)] J_R(\epsilon)J_L(\epsilon+\omega_0)
d\epsilon \label{eq:eqP} \eea
\end{widetext}
One could reason this rate equation as follows. In the first
equation, the term ${\mathcal P}_t(1,N,\omega)k_{1\rightarrow 0}^e$
stands for the decay rate of ${\mathcal P}_t(1,N,\omega)$. The
second line describes a process where by the time $t$ the TLS
occupies the ground state, $N-1$ excess electrons have arrived at
the $R$ terminal, and an overall of $\omega-\epsilon+\omega_0$
energy has been absorbed at the $R$ bath. At the time $t$ an
electron-hole pair excitation generates an electron at the $R$ bath,
leaving a hole at the $L$ metal. This charge transfer process is
accompanied by an electronic energy transmission at the amount of
$\epsilon-\omega_0$: An electron leaving the $L$ bath has a total
energy $\epsilon$, however only $\epsilon-\omega_0$ is gained by the
$R$ bath. The rest, at the amount of $\omega_0$, is gained by the
vibrational mode. A similar reasoning can explain other terms in
Eq. (\ref{eq:eqP}). For convenience, the factor $(2\pi)^{-1}$ in Eq. (\ref{eq:rate2})
has been absorbed into the definition
of $J_{\nu}(\omega)$. %XXXX

We Fourier transform the above system of equations with respect to
both charge and energy, to obtain the {\it characteristic function}
$\mathcal Z(\chi,\eta,t)$. It depends on the energy counting field $\eta$ and
the charge counting field $\chi$,
\bea
\ket{\mathcal{Z}(\chi,\eta,t)} \equiv
\begin{pmatrix}
\sum_{N=-\infty}^{\infty} e^{iN\chi}\intinf \mathcal P_t(0,N,\omega)e^{i\omega\eta}\,d\omega
\\ \sum_{N=-\infty}^{\infty} e^{iN\chi}\intinf
\mathcal P_t(1,N,\omega)e^{i\omega\eta}\,d\omega
\end{pmatrix}
\nonumber\\
\label{eq:z}
\eea
It satisfies the differential equation
\beq \d{\ket{\mathcal{Z}(\chi, \eta,t)}}{t} = - \mathcal{\hat W}(\chi,\eta)\ket{\mathcal Z(\chi,\eta, t)},
\label{eq:Z} \eeq
where the matrix $\mathcal{\hat W}$ contains the following elements
\bea &&\mathcal{\hat W}(\chi,\eta) =
\nonumber\\
&&\begin{pmatrix}
k_{0\rightarrow 1}^{L\rightarrow R} + k_{0\rightarrow 1}^{R\rightarrow L}  & -e^{i\chi}F_1^-(\eta)-e^{-i\chi}F_2^+(\eta)  \\
-e^{i\chi}F_1^+(\eta)-e^{-i\chi}F_2^-(\eta)  & k_{1\rightarrow 0}^{L\rightarrow R} + k_{1\rightarrow0}^{R\rightarrow L}   \\
\end{pmatrix}
\nonumber\\ \label{eq:mu} \eea
Here,
\bea &&F_1^{\pm}(\eta)=
\nonumber\\
&&\intinf e^{i\epsilon\eta}
f_L(\epsilon\pm\omega_0)[1-f_R(\epsilon)]J_L(\epsilon\pm\omega_0)J_R(\epsilon)d\epsilon
\nonumber\\
&&F_2^{\pm}(\eta)=
\nonumber\\
&&\intinf e^{-i\epsilon\eta}
[1-f_L(\epsilon\pm\omega_0)]f_R(\epsilon)J_L(\epsilon\pm\omega_0)J_R(\epsilon)d\epsilon
\nonumber\\
%F_3(\eta)=\intinf e^{i\epsilon\eta} f_L(\epsilon+\omega_0)[1-f_R(\epsilon)]J_L(\epsilon+\omega_0)J_R(\epsilon)d\epsilon
\nonumber\\
%F_4(\eta)=\intinf e^{-i\epsilon\eta} [1-f_L(\epsilon-\omega_0)]f_R(\epsilon)J_L(\epsilon-\omega_0)J_R(\epsilon)d\epsilon
\eea
The {\it cumulant generating function} is formally defined as
\bea
&&G(\chi,\eta)=
\nonumber\\
&&\lim_{t \to \infty} \ \frac{1}{t}\ln \sum_{N=-\infty}^{\infty} e^{iN\chi} \intinf  \mathcal
P_t(N,\omega)e^{i\omega\eta}d\omega, \label{eq:G} \eea
where we introduced the short notation ${\mathcal
P}_{t}(N,\omega)={\mathcal P}_t(0,N,\omega)+{\mathcal
P}_t(1,N,\omega)$, that is the probability to transfer by the time
$t$, $N$ electrons and an energy $\omega$ from left to right,
irrespective of the state of the TLS.
The charge and heat currents can be readily derived, by taking the
first derivative of the CGF with respect to either $\eta$ or $\chi$,
\bea
&&\langle I_e\rangle \equiv \frac{\avg{N}_{t}}{t} = \d{G(\chi,\eta)}{(i\chi)}\Big|_{\chi=0,\eta=0}
\nonumber\\
&&\langle I_q\rangle \equiv \frac{\avg{\omega}_{t}}{t} = \d{G(\chi,\eta)}{(i\eta)}\Big|_{\chi=0,\eta=0}
\label{eq:J}
\eea
The quantity $\avg{\omega}_{t}$ denotes the total energy $\omega$
transferred from $L$ to $R$ by the (infinitely long) time $t$;
$\avg{N}_t$ similarly counts the particles (electrons) transferred
in the same direction, by that time. The zero frequency noise
current power density can be similarly obtained,
\bea
&& \avg{S_e}  \equiv \frac{\avg{N^2}_{t} - \avg{N}^2_{t}}{t} =  \dd {G(\chi,\eta)}{(i\chi)}\Big|_{\chi=0,\eta=0}
\nonumber\\
&&\avg{S_q}  \equiv \frac{\avg{\omega^2}_{t} - \avg{\omega}^2_{t}}{t} =  \dd {G(\chi,\eta)}{(i\eta)}\Big|_{\chi=0,\eta=0}.
\label{eq:S}
\eea

The CGF can be expressed in terms of $|\mathcal Z\rangle$ as
\bea
G(\chi,\eta) =  \lim_{t \to \infty} \ \frac{1}{t}\ln \langle I| \mathcal Z(\chi,\eta,t) \rangle,
\eea
with $\langle I|= \langle 1 1|$,  a left vector of unity. It
is practically given by the negative of the smallest eigenvalue of
the matrix $\mathcal{\hat W}$,
\bea
G(\chi,\eta)&=& -\frac{w_{1,1} + w_{2,2}}{2}
\nonumber\\
& +& \frac{\sqrt{
(w_{1,1}-w_{2,2})^2+4w_{1,2}(\chi,\eta)w_{2,1}(\chi,\eta)}}{2}.
\nonumber\\ \label{eq:QE} \eea
$w_{i,j}$ are the matrix elements of $\mathcal{\hat W}$, see Eq.
(\ref{eq:mu}).

%================================================
\subsection{Fluctuation theorem}

We confirm next the following symmetry
\bea
G(\chi,\eta)= G(-\chi+i(\beta_L\mu_L-\beta_R\mu_R),-\eta+i\Delta\beta),
\label{eq:FT}
\eea
with $\Delta \beta=\beta_R-\beta_L$. In order to prove this,
we focus on the product $D(\chi,\eta)\equiv
w_{1,2}(\chi,\eta)w_{2,1}(\chi,\eta)$ in Eq. (\ref{eq:QE}),
\bea
D(\chi,\eta)&=&
\left[ e^{i\chi}F_1^{-}(\eta) + e^{-i\chi}F_2^+(\eta)\right]
\nonumber\\
&\times&
\left[ e^{i\chi}F_1^{+}(\eta) + e^{-i\chi}F_2^-(\eta) \right].
\eea
Under the transformation $\chi\rightarrow
-\chi+i(\beta_L\mu_L-\beta_R\mu_R)$ and $\eta \rightarrow
-\eta+i\Delta\beta$, using the relation
$f_{\nu}(\epsilon)=[1-f_{\nu}(\epsilon)]e^{-\beta_{\nu}(\epsilon-\mu_{\nu})}$,
we find that
\begin{widetext}
\bea e^{i\chi}F_1^-(\eta) &\rightarrow&
e^{-i\chi}e^{-\beta_L\mu_L+\beta_R\mu_R} \intinf d\epsilon
e^{-i\epsilon\eta}e^{-\Delta\beta\epsilon} \left[
1-f_L(\epsilon-\omega_0)
\right]e^{-\beta_L(\epsilon-\omega_0-\mu_L)}
f_R(\epsilon)e^{\beta_R(\epsilon-\mu_R)}
J_L(\epsilon-\omega_0)J_R(\epsilon)
\nonumber\\
&=& e^{-i\chi}e^{\beta_L\omega_0}F_2^-(\eta).
\nonumber\\
e^{-i\chi}F_2^+(\eta) &\rightarrow&
e^{i\chi}e^{\beta_L\mu_L-\beta_R\mu_R} \intinf d\epsilon
e^{i\epsilon\eta}e^{\Delta\beta\epsilon} \left[ 1-f_R(\epsilon)
\right]e^{-\beta_R(\epsilon-\mu_R)}
f_L(\epsilon+\omega_0)e^{\beta_L(\epsilon+\omega_0-\mu_L)}J_L(\epsilon+\omega_0)J_R(\epsilon)
\nonumber\\
&=& e^{i\chi}e^{\beta_L\omega_0}F_1^+(\eta) . \label{eq:fr1} \eea
\end{widetext}
Similarly, one could show that
\bea
e^{i\chi}F_1^+(\eta)\rightarrow e^{-i\chi e^{-\beta_L\omega_0}}F_2^+(\eta)
\nonumber\\
e^{-i\chi}F_2^-(\eta)\rightarrow e^{i\chi e^{-\beta_L\omega_0}}F_1^-(\eta).
 \label{eq:fr2}
\eea
The extra factors $e^{\pm\beta_L\omega_0}$ cancel, and we recover
the symmetry
\bea D(\chi,\eta)=
D(-\chi+i(\beta_L\mu_L-\beta_R\mu_R),-\eta+i\Delta\beta), \eea
confirming Eq. (\ref{eq:FT}). We can now demonstrate the validity of a
fluctuation relation for this non-equilibrium system. The
probability to transfer the energy $\omega$ by the long time $t$, from
$L$ to $R$, is given by the inverse Fourier transform of Eq.
(\ref{eq:G}),
\bea {\mathcal P}_{t}(N,\omega)
\sim \frac{1}{2\pi}\sum_{-\infty}^{\infty}
e^{-iN\chi}\int_{-\infty}^{\infty}
%\mathcal Z(\chi,\eta,t)
C(\chi,\eta)e^{G(\chi,\eta)t}
e^{-i\omega\eta} d\eta,
\nonumber\\
\label{eq:Pt} 
\eea
%
% CHECK inverse Fourier transf for discrete FT
with $ \lim_{t \to \infty} [\ln C(\chi,\eta)] /t=0$.
Similarly, the quantity ${\mathcal P}_{t}(-N,-\omega)$ represents
the probability that $N$ charged particles and an  energy $\omega$
have been transmitted in the opposite direction, right to left, up
to time $t$. Based on the symmetry Eq. (\ref{eq:FT}), one can show
that \cite{Mukamel-rev}
\bea \lim_{t \to \infty} \ \frac{1}{t} \ln \frac{{\mathcal
P}_{t}(N,\omega)}{{\mathcal P}_{t}(-N,-\omega)}=\frac{\omega \Delta\beta  + N(\beta_L\mu_L-\beta_R\mu_R)  }{t},
\label{eq:SSFTM} \eea
which is often written in a compact form as
\bea
\frac{{\mathcal P}_{t}(N,\omega)}{{\mathcal P}_{t}(-N,-\omega)}=e^{\omega \Delta\beta  + N(\beta_L\mu_L-\beta_R\mu_R)}.\label{eq:SSFTMs}
\eea
This expression goes beyond standard metal-molecule weak-coupling
schemes as the energy and charge transfer and not tightly coupled,
and the energy $\omega$ can take continuous values, unlike Refs.
\cite{Mukamel-weak,HanggiBerry,Gernot}.

It should be noted that the above derivation has assumed charge and
energy conservation between the two reservoirs. The full
particle-energy counting statistics, without such an assumption,
would begin with the probability distribution $\mathcal
P_t(n,N_L,N_R,\omega_L,\omega_R)$, to find the system at time $t$ in
the spin state $n=0,1$, with $N_{\nu}$ electrons and $\omega_{\nu}$
excess energy accumulated at the $\nu$ bath. One can readily write
an equation of motion for this function, analogous to Eq.
(\ref{eq:eqP}), to be Fourier transformed using four counting fields,
\bea
&&\mathcal P_t(n,\chi_L,\chi_R,\eta_L,\eta_R)
=\sum_{N_L}e^{iN_L\chi_L}\sum_{N_R} e^{iN_R\chi_R}
\nonumber\\
&&\times\intinf e^{i\omega_L\eta_L}d\omega_L\intinf e^{i\omega_R\eta_R}
\mathcal P_t(n,N_L,N_R,\omega_L,\omega_R) d\omega_R.
\nonumber\\
\eea
This quantity satisfies an equation of motion that is analogous to
Eq. (\ref{eq:Z}). It can be readily proved that the negative of the
smallest eigenvalue of the corresponding matrix $\mathcal{\hat
W}(\chi_L,\chi_R,\eta_L,\eta_R)$ obeys the symmetry
\bea
&&G(\chi_L,\chi_R,\eta_L,\eta_R) =
\nonumber\\
&&
G(-\chi_L+i\beta_L\mu_L, -\chi_R+i\beta_R\mu_R, -\eta_L+i\beta_L,-\eta_R+i\beta_R),
\nonumber\\
\eea
which can be translated into the FT for the probability itself,
\bea
\frac{{\mathcal P}_{t}(N_L,N_R,\omega_L,\omega_R)}
{{\mathcal P}_{t}(-N_L,-N_R,-\omega_L,-\omega_R) }
&=&e^{(N_L\beta_L\mu_L+N_R\beta_R\mu_R)}
\nonumber\\
&\times& e^{(\beta_R\omega_R+\beta_L\omega_L)}. \eea
Here, $\mathcal
P_t(N_L,N_R,\omega_L,\omega_R)=$$\mathcal
\sum_{n=0,1}\mathcal P_t(n,N_L,N_R,\omega_L,\omega_R)$.
Enforcing energy and charge conservation, $N=N_L=-N_R$ and
$\omega=\omega_R=-\omega_L$, we recover Eq. (\ref{eq:SSFTMs}).

%=============================
\subsection{Currents, and measures for vibrational cooling, heating, or instability}

{\it Currents.} Analytical expressions for the charge and energy
currents are obtained using the definition Eq. (\ref{eq:J}),
utilizing Eqs. (\ref{eq:mu}) and (\ref{eq:QE}). These currents are
defined positive when flowing $L$ to $R$, and their closed forms are
\bea \avg{I_e}=p_1 (k_{1\rightarrow0}^{L\rightarrow R} -
k_{1\rightarrow0}^{R\rightarrow L})
 +p_0 (k_{0\rightarrow1}^{L\rightarrow R} -  k_{0\rightarrow1}^{R\rightarrow L}),
\label{eq:Ie}
\eea
and
\bea
&&\avg{I_q}=
\nonumber\\
&&p_1 \Big[ \intinf d\omega \omega f_L(\omega-\omega_0)[1-f_R(\omega)]J_L(\omega-\omega_0)J_R(\omega)
\nonumber\\
&&-\intinf d\omega \omega [1-f_L(\omega+\omega_0)]f_R(\omega)J_L(\omega+\omega_0)J_R(\omega) \Big]
\nonumber\\
&&+p_0 \Big[ \intinf d\omega \omega f_L(\omega+\omega_0)[1-f_R(\omega)]J_L(\omega+\omega_0)J_R(\omega)
\nonumber\\
&&-\intinf d\omega \omega
[1-f_L(\omega-\omega_0)]f_R(\omega)J_L(\omega-\omega_0)J_R(\omega)
\Big]. \nonumber\\
\label{eq:Iq}
\eea
The TLS population is calculated in the steady-state limit,
\bea
p_1=\frac{k_{0\rightarrow 1}^e}{k_{0\rightarrow 1}^e +k_{1\rightarrow 0}^e }; \,\,\,\, p_0=1-p_1.
\eea
The zero frequency noise current power is given by
\bea
&&\avg{S_e}= -\frac{2}{k_{0\rightarrow 1}^e+k_{1\rightarrow 0}^e} \avg{I_e}^2
\nonumber\\
&&+ \frac{4}{k_{0\rightarrow 1}^e+k_{1\rightarrow 0}^e}
(k_{0\rightarrow 1}^{L\rightarrow R} k_{1\rightarrow 0}^{L\rightarrow R} +
k_{0\rightarrow 1}^{R\rightarrow L} k_{1\rightarrow 0}^{R\rightarrow L}).
\eea
The energy current, directed towards the vibrational mode, is zero
in the steady-state limit, unless the mode is further coupled to a
dissipative bath. Formally, it is given by the expression
\bea \avg{I_{vib}}=-\omega_0p_1 \left[ k_{1\rightarrow
0}^{L\rightarrow R} + k_{1\rightarrow 0}^{R\rightarrow L} \right]
%\nonumber\\
%&+&\intinf d\omega  [1-f_L(\omega+\omega_0)]f_R(\omega)J_L(\omega+\omega_0)J_R(\omega) \Big]
+p_0 \omega_0 \left[  k_{0\rightarrow 1}^{L\rightarrow R}
+k_{0\rightarrow 1}^{R\rightarrow L} \right].
%\Big[ \intinf d\omega
%f_L(\omega+\omega_0)[1-f_R(\omega)]J_L(\omega+\omega_0)J_R(\omega)
%\nonumber\\
%&+&\intinf d\omega [1-f_L(\omega-\omega_0)]f_R(\omega)J_L(\omega-\omega_0)J_R(\omega) \Big]
\nonumber\\
\label{eq:Iph}
\eea

{\it Measures for vibrational instability.} The stability of the
junction can be estimated, against heating effects, by inspecting
several measures. First, following Ref. \cite{Lu}, we define the
{\it damping rate} $K_{vib}$ of the vibrational mode as the
difference between relaxation and excitation rates,
\bea K_{vib}\equiv k_{1\rightarrow 0}^e-k_{0\rightarrow1}^e.
\label{eq:Kvib} \eea
Positive $K_{vib}$ indicates on the ``normal" thermal-like behavior,
as relaxation processes overcome excitations. In this case, the mode
effective temperature (defined below) is found to be either below (cooling)
or above (heating) the environmental temperature, yet the junction
remains stable in the sense that the ground vibrational state population is
larger than the excited level population. A negative value for $K_{vib}$
evinces on the process of an {\it uncontrolled} heating of the
molecular mode, eventually leading to junction instability and
breakdown. One can also directly inspect the TLS population:
population inversion reflects on vibrational instability.

{\it Effective temperature.} The TLS population can be further
utilized as a measure for the  molecular vibration
effective temperature, $1/\beta_{eff}$, {\it defined} using an
equilibrium relation,
\bea \frac{p_1}{p_0}=e^{-\beta_{eff}\omega_0}. \label{eq:Teff} \eea
A negative value for $\beta_{eff}$ attests on population inversion,
thus junction instability. When $\beta_{eff}$ is positive, one
should compare it to the reservoirs' inverse temperature $\beta$: If
$\beta_{eff}>\beta$ the system demonstrates bias-induced cooling
phenomena. For $\beta_{eff}<\beta$ the vibrational mode is heated
up relative to its environment.
The latter typically occurs at an intermediate bias voltage,
before instabilities take place.

%===========================
%======================================
% Current
\begin{figure}[htbp]
\vspace{1mm}  {\hbox{\epsfxsize=75mm
\hspace{00mm}\epsffile{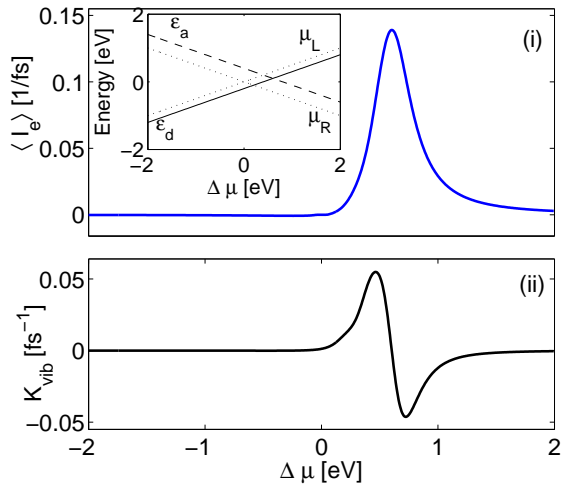}}}
\caption{(i) Charge current in a rectifying molecular junction.
Inset: Energies of the donor (full line) and acceptor states (dashed line).
The dotted lines correspond to the chemical potential at the left and right sides.
(ii) Damping rate $K_{vib}$.
The junction parameters are
$\Gamma_{\nu}$=0.2, $1/\beta_{\nu}=0.005$, $\kappa=0.1$, $\omega_0=0.05$ and
$\epsilon_d(\Delta\mu=0)=-0.2$,
$\epsilon_a(\Delta\mu=0)=0.4$, all in units of [eV].  } \label{Fig1}
\end{figure}

% pop
\begin{figure}[htbp]
\vspace{1mm}  {\hbox{\epsfxsize=75mm \hspace{00mm}\epsffile{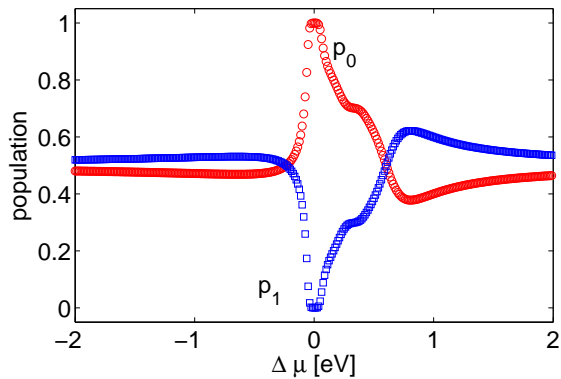}}}
\caption{Population of the two-state ``vibration" as a function of bias
voltage. Parameters are the same as in Fig. \ref{Fig1}.}
\label{Fig1p}
\end{figure}

%======================================

\subsection{Numerical results: isolated mode}

We demonstrate cooling, heating and mode instability
upon varying the bias voltage.
A generic mechanism leading to vibrational instabilities (and eventually junction rupture)
in D-A  molecular rectifiers has been discussed
in Ref. \cite{Lu}: At large positive bias, when
the D state is positioned above the acceptor level,
electron-hole pair excitations by the molecular
vibration (TLS here) dominate the mode dynamics.
This can be schematically seen in Fig. \ref{FigR}, where the rate $k_{0\rightarrow1}^{L\rightarrow R}$
overcomes other rates once the donor spectral function is positioned above the acceptor spectral
function. As $K_{vib}$ becomes negative, 
% leading to ``negative" damping of energy, i.e., to 
population inversion is observed.

The junction setup is displayed in Fig. \ref{Fig1}. D and A
levels are positioned such that in equilibrium, $\Delta \mu=0$, the
donor level is placed below the Fermi energy $\mu$, while the
acceptor level is of a higher energy, $\epsilon_d(\Delta
\mu=0)<\mu<\epsilon_a(\Delta \mu=0)$. Under an applied bias, the
levels are assumed to linearly follow the external potential drive
(inset) \cite{bias}. Therefore, at a particular positive bias the
levels cross. Beyond that, the levels exchange arrangement, and the D state
is of a higher energy. Throughout the paper, the parameters
$\omega_0$, $\Gamma_{\nu}$, $\Gamma_{ph}$, $1/\beta$, $\kappa$,
$\epsilon_{d,a}$ and $\Delta\mu$ are given in units of eV.

The junction's current-voltage characteristics is displayed in Fig.
\ref{Fig1} (i), manifesting a substantial rectification effect. For
negative polarity, $\Delta \mu=\mu_L-\mu_R<0$, the current is rather
small. In contrast, for positive bias the current substantially
increases once $\Delta\mu>\omega_0$, reaching a maximum when the
energy levels satisfy $\epsilon_d-\epsilon_a\sim\omega_0$. Level
broadening, $\Gamma_{\nu}$, affects the actual position of the
maximum.
The damping rate, $K_{vib}$, is displayed in Fig. \ref{Fig1}(ii). It
shows the following features: First, for large negative bias,
$\Delta\mu<-0.2$,  $K_{vib}$ is negative. This instability can be
immediately removed, once a very weak coupling to a phononic
 thermal reservoir is turned on, see Figs. \ref{Fig4} and \ref{FigS1} below.
Beyond that, the damping rate $K_{vib}$ is positive between
$-0.2\lesssim\Delta\mu\lesssim0.6$, indicating on a stable mode of
operation. However, for large enough bias, $\Delta\mu\gtrsim 0.6$,
once $\epsilon_d>\epsilon_a$, uncontrolled TLS heating takes place,
recognized by a sign change in  $K_{vib}$. It should be noted that
the instability takes place in the parameter range very relevant to the
rectifier operation. It is thus important to understand how to
tune the system configuration so as to sustain junction functionality.

Fig. \ref{Fig1p} depicts the corresponding population of the two
levels. At zero bias, $k_{0\rightarrow 1}^e=0$, thus the population of
the excited state is identically zero. At low positive bias one
finds that $k_{0\rightarrow 1}^e< k_{1\rightarrow 0}^e$, leading to the
``normal" situation of $p_0>p_1$. However, once the bias is large
and the donor state is positioned above the acceptor site, ($\Delta
\mu \sim0.6$) the excitation rate $k_{0\rightarrow 1}^e$ exceeds the
relaxation rate $k_{1\rightarrow 0}^e$ and population inversion takes
place. We note that for a negative bias, small population inversion
is also observed, as electrons damp energy to the TLS when crossing
the junction. However, since $\avg {I_e}$ is rather small 
(Fig. \ref{Fig1}), we do not expect molecular instability in
this regime, see also Fig. \ref{FigS1}.

\begin{figure}[htbp]
\vspace{1mm}  {\hbox{\epsfxsize=75mm \hspace{0mm}\epsffile{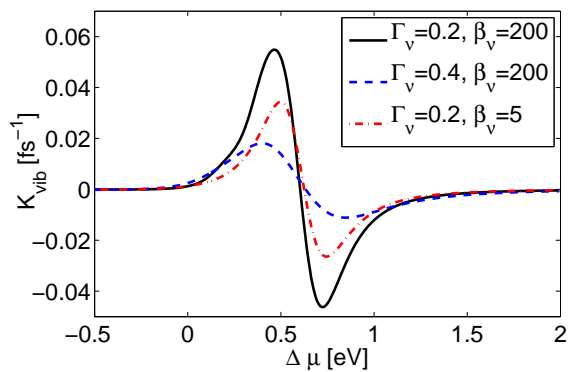}}}
\caption{ Damping rate in a rectifying junction for different
broadening parameters, $\Gamma_{\nu}$=0.2, $\beta_{\nu}=200$  (full)
$\Gamma_{\nu}$=0.4, $\beta_{\nu}=200$  (dashed) $\Gamma_{\nu}$=0.2,
$\beta_{\nu}=5$ (dashed-dotted). Other parameters are the same as in
Fig. \ref{Fig1}. } \label{Fig2}
\end{figure}

The details of the damping rate $K_{vib}$ depend on the level
broadening and the reservoirs temperature as we show in Fig.
\ref{Fig2}. The position of the turnover, between positive to
negative damping, appears at a similar value for the bias, and it is
generally independent of the reservoirs temperatures and
$\Gamma_{\nu}$. However, the width of the curve largely depends on
these parameters.

It should be noted that the development of the instability, as reported in 
Figs. \ref{Fig1}, \ref{Fig1p} and \ref{Fig2},
{\it does not} depend on the concrete value of $\kappa$, the strength of the molecule-mode coupling, and
the behavior persists in the limit of vanishing vibronic coupling, $\kappa\rightarrow 0$.
In the next section we allow the vibrational mode to thermalize with a 
phononic environment at a rate $\Gamma_{ph}$.
In this case, the competition between $\kappa$ and $\Gamma_{ph}$ determines the onset of instability,
see Eq. (\ref{eq:Kvib2}).

% K
%======================================

%======================================

\subsection{Numerical results: dissipative mode}

Up to this point, we have assumed that 
%electronic transitions are
%coupled to a 
the molecular vibrational mode (TLS here) is well
isolated from other vibrations. In reality, internal modes typically
exchange energy with  ``secondary" reservoirs modes, either
internal, or part of a larger environment, opening up an additional
route for energy dissipation. It is expected that in the presence of
such a thermal bath, the region of vibrational instability
($K_{vib}<0$) would become limited.

A simple model that is capable of describing a hierarchy of energy
transfer processes, electronic energy $\rightarrow$ specific
vibrational excitation $\rightarrow$ thermal bath, is given by an
extension of the model (\ref{eq:HBB}),
\bea H_{A+B} &=& \frac{\omega_0}{2} \sigma_z + \sigma_x
\left(F_e+F_b\right)
\nonumber\\
&+& \sum_l  \epsilon_l
a_l^{\dagger}a_l + \sum_r  \epsilon_r
a_r^{\dagger}a_r +\sum_{\alpha}\omega_{\alpha}b_{\alpha}^{\dagger}b_{\alpha}.
\label{eq:HBBB}
\eea
The notation ``$H_{A+B}$" indicates that the anharmonic mode is coupled to
a thermal bath (B).
The operator $F_e$ describes electron-hole pair excitations as in Eq.
(\ref{eq:Fe}).
The thermal bath operator, coupled to the TLS transitions,
includes displacements of reservoir modes,
\bea F_b=\sum_{\alpha} v_{\alpha}(b_{\alpha}^{\dagger}+b_{\alpha}),
\label{eq:Fb} \eea
with $b_{\alpha}^{\dagger}$ ($b_{\alpha}$) as a bosonic creation (annihilation)
operator for the $\alpha$ phonon-reservoir mode.

Derivation of the full counting statistics can be reiterated, while
including energy dissipation from the TLS to the phonon bath. For
details, see Appendix A. We find that the expression for the charge
current stays intact, satisfying the formal expression
(\ref{eq:Ie}). However, the steady-state populations are corrected
by a phonon relaxation rate constant as
\bea p_1=\frac{ k_{0\rightarrow1}^e + \Gamma_{ph}(\omega_0)
n_{ph}(\omega_0)} { k_{0\rightarrow1}^e+k_{1\rightarrow0}^e+
\Gamma_{ph}(\omega_0)[2n_{ph}(\omega_0)+1] }. \eea
The electronic transition induced rates $k_{n\rightarrow n'}^e$ are
those defined in Eq. (\ref{eq:kTLS}); the phononic relaxation rate constant
is $\Gamma_{ph}(\omega)=2\pi\sum_{\alpha}v_{\alpha}^2
\delta(\omega_{\alpha}-\omega)$. The function
$n_{ph}(\omega)=[e^{\beta_{ph}\omega}-1]^{-1}$ stands for the
Bose-Einstein distribution with $\beta_{ph}$ 
as the temperature of the phonon bath.

\begin{figure}[htbp]
\vspace{1mm}  {\hbox{\epsfxsize=70mm \hspace{0mm}\epsffile{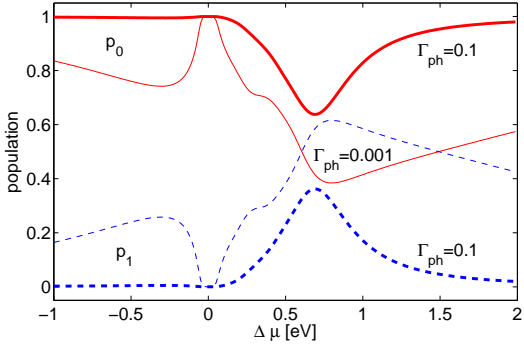}}}
\caption{TLS population as a function of bias voltage for
$\Gamma_{ph}=0.001$  (narrow lines) and  $\Gamma_{ph}=0.1$  (heavy
lines). The excited (ground) state population is presented by dashed
(full) lines. Other parameters are the same as in Fig. \ref{Fig1}
with $\beta_{ph}=200$. } \label{Fig4}
\end{figure}

Fig. \ref{Fig4} presents the steady-state population for two choices
of $\Gamma_{ph}$. When this
parameter is small, population inversion still takes place around
donor-acceptor level crossing. However, the phenomenon disappears at
{\it large enough bias}. Thus, quite interestingly, the domain of
instability extends {\it intermediate} bias values, while the system
becomes stable again at very high bias. This can be reasoned by
inspecting $K_{vib}$. It is  defined as the difference between TLS
relaxation and excitation rate constants. In the presence of a
thermal bath it is given by
\bea
K_{vib}&=&\left(k_{1\rightarrow 0}^e + \Gamma_{ph}(n_{ph}+1) \right)
-\left( k_{0\rightarrow 1}^e+\Gamma_{ph} n_{ph} \right).
\nonumber\\
&=& k_{1\rightarrow 0}^e - k_{0\rightarrow 1}^e + \Gamma_{ph}.
\label{eq:Kvib2}
\eea
For convenience, the $\omega_0$ dependence of the rates is left out.
While $k_{1\rightarrow 0}^e < k_{0\rightarrow 1}^e$ may hold at
large bias, both these rates diminish with $\Delta \mu$, and the net
damping rate can become positive due to the $\Gamma_{ph}$
contribution. For large enough $\Gamma_{ph}$, instability does not
take place at any voltage.

%==================================================
% pop with bath

% TLS temp
\begin{figure}[htbp]
\vspace{1mm}  {\hbox{\epsfxsize=75mm\hspace{0mm}\epsffile{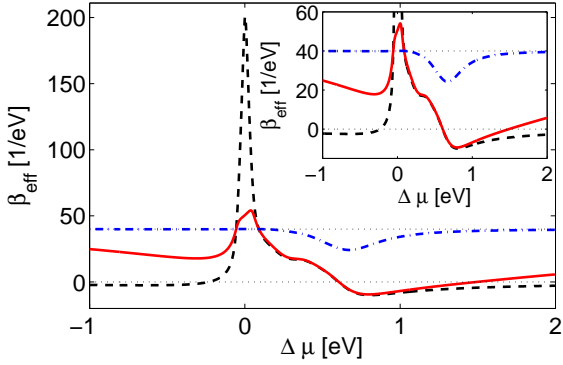}}}
\caption{
Effective TLS temperature,
$\Gamma_{ph}=0$ (dashed line); $\Gamma_{ph}=0.001$ and $\beta_{ph}=40$ (full line)
and  $\Gamma_{ph}=0.4$ and $\beta_{ph}=40$ (dashed-dotted line).
The inset zooms on the latter two cases.
The dotted lines mark the values $\beta_{ph}=40$ and $\beta=0$.
Other junction parameters are the same as in Fig. \ref{Fig1},
 with $\beta_{\nu}=200$.
}
\label{Figb}
\end{figure}

% cooling
\begin{figure}[htbp]
\vspace{1mm}  {\hbox{\epsfxsize=70mm\hspace{0mm}\epsffile{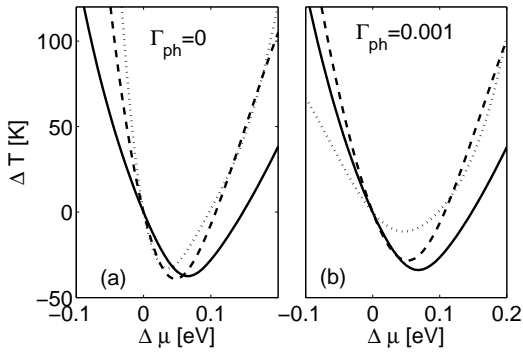}}}
\caption{
Cooling of the molecular vibration
for $\omega_0=0.05$ (dotted line), $\omega_0=0.15$ (dashed line),
$\omega_0=0.3$ (full line).
(a): $\Gamma_{ph}=0$.  (b): $\Gamma_{ph}=0.001$.
Other junction parameters are  $\Gamma_L=\Gamma_R=0.1$,
and $\beta_{ph}=\beta_{\nu}=40$.
%In the absence of bias $\epsilon_d=-0.2$ and $\epsilon_a=0.4$ .
The levels are shifted with the bias voltage as depicted in Fig. \ref{Fig1}.
}
\label{Figb2}
\end{figure}

%---------------------------------------

The effective TLS temperature, defined in Eq. (\ref{eq:Teff}), is
displayed in Fig. \ref{Figb} for several cases. First, in the
absence of a phonon thermal bath we find that at zero bias voltage
the molecular mode is thermalized at the metals' temperature,
$\beta_{eff}=\beta$. This effective inverse temperature quickly
drops with increasing bias, becoming negative around the value of
$\Delta \mu=0.6$, where the D-A levels cross. This behavior
indicates on the instability of the junction from that point. Next,
we weakly couple ($\Gamma_{ph}$=0.001) the single mode to an
additional thermal bath maintained at $\beta_{ph}=40$. The following
observations can be made: (i) For negative bias, the mode is close
to be equilibrated with the phonon bath, as electron-hole
excitations are sparse. (ii) In accordance with Fig. \ref{Fig4},
$\beta_{eff}$ can reach a negative (unstable) value around
$\Delta\mu=0.6$. However, $\beta_{eff}$ becomes positive at large
enough bias, indicating that the system re-enter a stability region.
(iii) At low bias, $-0.05<\Delta \mu<0.1$, the mode temperature is
lower than its phononic environment, as $\beta_{eff}>\beta_{ph}$.
(iv) At strong mode-thermal bath coupling, $\Gamma_{ph}=0.4$,  the
mode is closed to be thermalized with $\beta_{ph}$ at all biases.

We now demonstrate mode cooling, to a temperature below the phonon
bath and metals temperature. Keeping both electron and phonon
reservoirs at a fixed temperature of $\beta=40$, the temperature
difference $\Delta T\equiv T_{eff}-T_{ph}$ is presented in Fig.
\ref{Figb2}, for various frequencies and $\Gamma_{ph}$ values.
Generally, we note that at low positive bias, $\Delta \mu <0.1$, one
may cool the mode by 40 K, the result of its coupling to a
nonequilibrium environment.

%==================================================================

\section{Harmonic-mode Rectifier}

We study next the dynamics of model (\ref{eq:HTA}), assuming a
harmonic mode coupled to the electronic system. The relevant equations of
motion for the mode levels population are \cite{SegalM}
\bea
\dot p_n&=&-[nk_{d}^e + (n+1)k_{u}^e] p_n
\nonumber\\
&+&(n+1)k_{d}^ep_{n+1} + nk_u^ep_{n-1}.
\label{eq:DH}
\eea
Here, the decay rate constant is independent of the level index
$k_d^e\equiv k_{1\rightarrow 0}^e$, and similarly, $k_u^e\equiv
k_{0\rightarrow 1}^e$, defined in Eq. (\ref{eq:kTLS}). In order to
calculate the CGF, we define $\mathcal
P_t(n,N,\omega)$ as the probability that by the time $t$ the
harmonic mode occupies level $n$,  $N$ electrons have been added to
the $R$ bath and an additional energy $\omega$ has been acquired by
the $R$ bath. This quantity follows a differential equation
analogous to Eq. (\ref{eq:eqP}). The characteristic function is an
array whose $n$th element is $ |\mathcal
Z(\chi,\eta,t)\rangle_n =\sum_N e^{iN\chi}\intinf \mathcal
P_t(n,N,\omega)e^{i\omega\eta}\,d\omega$. It satisfies a
differential equation corresponding to Eq. (\ref{eq:Z})
\beq \d{\ket{\mathcal Z(\chi, \eta,t)}}{t} = - \mathcal{\hat {W}}(\chi,\eta)\ket{ \mathcal Z(\chi,\eta, t)}, \label{eq:ZH} \eeq
with the $n\times n$  matrix $\mathcal {\hat W(\chi,\eta)}$,
\begin{widetext} \bea &&\mathcal{\hat {W}}(\chi,\eta) =
\nonumber\\
&&\begin{pmatrix}
k_{u}^e  & -e^{i\chi}F_1^-(\eta)-e^{-i\chi}F_2^+(\eta)  & 0 & 0 & ... &... \\
-e^{i\chi}F_1^+(\eta)-e^{-i\chi}F_2^-(\eta)  & k_{d}^e+2k_{u}^e &  -2e^{i\chi}F_1^-(\eta)-2e^{-i\chi}F_2^+(\eta)& 0 & ... &...  \\
0 & -2e^{i\chi}F_1^+(\eta)-2e^{-i\chi}F_2^-(\eta) & 2k_{d}^e+3k_{u}^e &-
3e^{i\chi}F_1^-(\eta)-3e^{-i\chi}F_2^+(\eta) &0 &...\\
0 &0&...&...&...&...&\\
...&...&\\
\end{pmatrix}
\nonumber\\ \label{eq:muH} \eea
\end{widetext}
We can readily confirm the fluctuation theorem, by inspecting the
eigenvalues of $\det[\lambda I -\mathcal{\hat W}]$. For
convenience, we define the auxiliary matrix $A \equiv\lambda I-\mathcal{\hat W}$.
Since it is tridiagonal, its determinant can be evaluated in a
recursive manner as
\bea \det [A]_{1,...,n}&=& a_{n,n}\det[A]_{1,...,n-1}
\nonumber\\
&-&a_{n,n-1}a_{n-1,n}\det[A]_{1,...,n-2} \eea
where $[A]_{1,...,k}$ denotes the submatrix constructed by the first
$k$ rows and columns of $A$. Thus, the symmetry of $\det[A]$ with respect
to $\chi$ and $\eta$ is determined by the symmetry of the
products $a_{n,n-1}a_{n-1,n}=w_{n,n-1}w_{n-1,n}$, with $w_{i,j}$ the matrix elements of 
$\mathcal {\hat W}$,
\bea &&d_n(\chi,\eta)\equiv w_{n,n-1}(\chi,\eta)w_{n-1,n}(\chi,\eta) \propto \nonumber\\
&&\left[ e^{i\chi}F_1^-(\eta)+e^{-i\chi}F_2^+(\eta)\right]
\left[ e^{i\chi}F_1^+(\eta)+e^{-i\chi}F_2^-(\eta) \right].
\nonumber\\
\eea
Using the relations (\ref{eq:fr1})-(\ref{eq:fr2}), we conclude that
\bea d_n(\chi,\eta)=
d_n(-\chi+i(\beta_L\mu_L-\beta_R\mu_R),-\eta+i\Delta\beta). \eea
Given the recursive nature of $\det[A]$, this
symmetry holds for all the eigenvalues of $\mathcal{\hat W}$,
confirming the fluctuation theorem (\ref{eq:FT}).
We now obtain the steady-state population of the harmonic mode, by
solving Eq. (\ref{eq:DH}) in the long time limit, $\dot p_n=0$. This
results in \cite{SegalM}
\bea p_n=\left(\frac{k_u^e}{k_d^e}\right)^n\frac{1}{\sum_{n=0}^{\infty}
(k_u^e/k_d^e)^n} ; \,\,\,\,\,\,\,\,\,\ n=0,1...,\infty\eea or \bea
p_n=\left(\frac{k_u^e}{k_d^e}\right)^n\left(1-\frac{k_u^e}{k_d^e}\right),
\label{eq:P} \eea
if $k_u^e<k_d^e$. In the opposite limit, the system passes into the
unstable regime, and the levels' population diverges. In that sense,
the harmonic model is unphysical as the number of states is not
bounded. One way to pull the system back into physical realm is to
couple the vibrational mode with a thermal bath, see
Appendix A. As explained above for the TLS-mode case, the following damping rate
is a measure for the junction stability,
\bea K_{vib}=k_d^e-k_u^e. \eea
A negative value indicates on junction
instability, as uncontrolled heating of the mode takes place. Using
steady-state populations, we proceed and derive the charge current
expression, valid only if $k_u^e<k_d^e$,
\bea
\avg{I_e} &=&
\frac{1}{t}\bra{1 1 \dots}\pd{}{i\chi}e^{-\mathcal{\hat W(\chi,\eta)}t}|_{\chi=0,\eta=0}
\ket{\mathcal{Z}(\chi,\eta,t=0)}
\nonumber\\
 &=&-\bra{1 1\dots} \pd{\mathcal{\hat {W}}}{i\chi}|_{\chi=0,\eta=0}\ket{P_{ss}} 
\nonumber\\
&=& - \left[ \frac{(k_d^{R\rightarrow L} - k_d^{L\rightarrow R})k_u^e }{k_d^e - k_u^e} 
+ \frac{(k_u^{R\rightarrow L}-k_u^{L\rightarrow R} )k_d^e }{k_d^e - k_u^e} \right] 
\nonumber\\ &=& -2\frac{k_u^{R\rightarrow L}k_d^{R\rightarrow L} -
k_u^{L\rightarrow R}k_d^{L\rightarrow R} }{k_d^e - k_u^e}
\label{eq:IeH}
 \eea
Here, $P_{ss}$ is a vector of steady-state population given by Eq. (\ref{eq:P}).
Equations (\ref{eq:DH})-(\ref{eq:IeH}) can be generalized to include the 
interaction of the harmonic mode
with a dissipative-thermal phonon bath.
Appendix A exemplifies this procedure for the anharmonic-mode model. 
In practice, the electronic induced
rates $k_d^e$ and  $k_u^e$ in Eq. (\ref{eq:P}) are augmented by a phononic contribution,
 $\Gamma_{ph}(\omega_0)[n_{ph}(\omega_0)+1]$ and $\Gamma_{ph}(\omega_0)n_{ph}(\omega_0)$,
respectively.

%=========================================

\begin{figure}[htbp]
\vspace{1mm}  {\hbox{\epsfxsize=70mm
\hspace{0mm}\epsffile{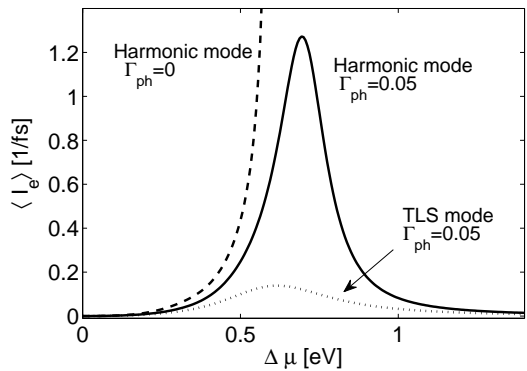}}}
\caption{
Charge current in the harmonic-mode model for $\Gamma_{ph}=0$ (dashed line) and
$\Gamma_{ph}=0.05$ (full line). For comparison, we also present
the current in the TLS-mode model with
$\Gamma_{ph}=0.05$ (dotted line).
$\beta_{\nu}=\beta_{ph}=40$,
other parameters are the same as in Fig. \ref{Fig1}. }
\label{FigH1}
\end{figure}

Fig. \ref{FigH1} displays the charge current for zero and finite $\Gamma_{ph}$ strength.
In the absence of coupling to the phonon bath, the current diverges around $\Delta \mu=0.6$,
where instability occurs.
For finite $\Gamma_{ph}$,  the current is larger when the vibrational mode is
harmonic, compared to the TLS-mode case (dotted),
as the electronic energy can be used to excite multiple transitions.

\begin{figure}[htbp]
\vspace{1mm}  {\hbox{\epsfxsize=75mm
\hspace{0mm}\epsffile{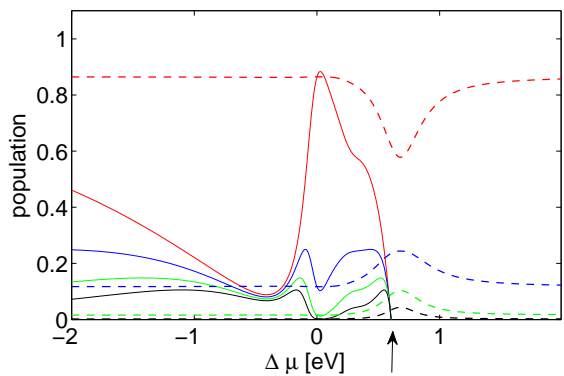}}}
\caption{
Population of the first four levels of the
harmonic mode.
Full lines: weak interaction with the heat bath, $\Gamma_{ph}$= $10^{-4}$.
The population becomes unphysical (negative) for $\Delta\mu>0.6$.
Dashed lines: strong interaction with heat bath, $\Gamma_{ph}$ = 0.1,
lifts the instability.
$\beta_{\nu}=\beta_{ph}=40$,
other parameters are the same as in Fig. \ref{Fig1}. }
\label{FigH2}
\end{figure}

Fig. \ref{FigH2} further demonstrates the ``stabilizing" effect the
interaction with a heat bath has on the harmonic mode. We
 display the population for the states $n=0$ to $n=3$, top
to bottom. % demonstrating a rich behavior: 
For $\Gamma_{ph}=0$  the
data is presented up to $\Delta \mu=0.6$, where the population
becomes unphysical (levels' population goes to zero there since
infinite number of vibrational states are occupied). This point is
indicated by an arrow.

%%=====================================================

%=========================================

The cooling and heating behavior depicted in Fig. \ref{Figb2} for
the TLS model could be repeated for the present harmonic-mode case
as well, to yield the  same behavior. The reason is that
$\beta_{eff}$ is determined by the ratio of rates,
and this ratio is identical in the two models. 
Our conclusion here is thus that
including mode-anharmonicity is important for transport
calculations: Harmonic-mode model can lead to unphysical results
(e.g., current divergence) since there is no saturation situation
for the vibrational mode. In particular, including molecular
anharmonic aspects is essential for obtaining reliable results when
simulating junction behavior close to the critical bias, where an
instability occurs.

%=========================================
\section{Summary}

We have studied vibrational cooling, heating, and instability formation in a
phonon assisted D-A electron rectifier junction using a
full-counting statistics approach. Variants of the basic model were
constructed, assuming either harmonic or an anharmonic vibrational
mode, further allowing energy dissipation to a phononic thermal environment.

\begin{figure}[htbp]
\vspace{1mm}  {\hbox{\epsfxsize=75mm \hspace{0mm}\epsffile{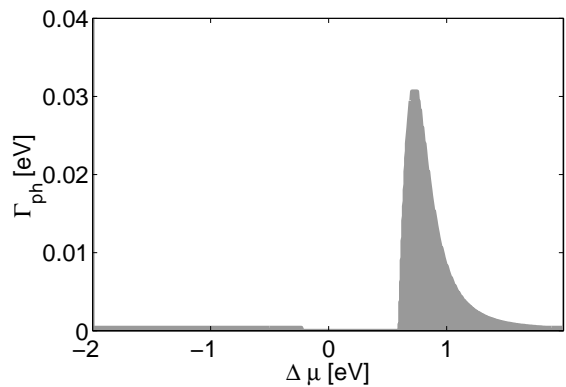}}}
\caption{Stability diagram. The dark island and the narrow strip
(at negative bias) are the parametric region in which the junction
becomes unstable. Other parameters are the same as in Fig.
\ref{Fig1}, besides the temperatures,
$\beta_L=\beta_R=\beta_{ph}=40$.} \label{FigS1}
\end{figure}

\begin{figure}[htbp]
\vspace{1mm}  {\hbox{\epsfxsize=75mm \hspace{0mm}\epsffile{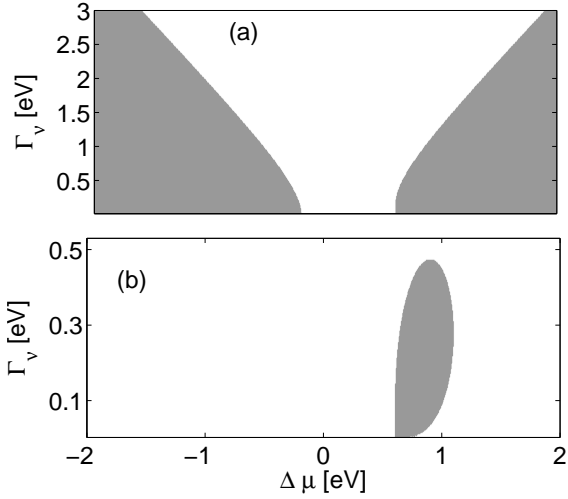}}}
\caption{(a): Stability diagram for $\Gamma_{ph}=0$. (b): Stability
diagram with $\Gamma_{ph}=0.005$. The dark region is the parametric
region in which the junction becomes unstable, $K_{vib}<0$. Other
parameters are the same as in Fig. \ref{Fig1}, besides the
temperatures $\beta_L=\beta_R=\beta_{ph}=40$.} \label{FigS2}
\end{figure}

Putting together our observations, we present in figures \ref{FigS1}
and \ref{FigS2} stability maps for the system; the dark region
codes instability zones, with negative $K_{vib}$. These diagrams hold
for both TLS and harmonic mode cases.
%
%In Fig. \ref{FigS1}, the black island at positive bias and the
%narrow strip at negative bias indicate on vibrational instability
%for small $\Gamma_ph$. Everywhere else, i.e., for strong enough
%mode-thermal bath interaction, the instability is lifted.
A reentrant behavior is observed in  Fig. \ref{FigS1}:
For a fixed value of $\Gamma_{ph}$, say
$\Gamma_{ph}=0.005$, the junction is stable for $\Delta\mu<0.6$,
unstable around $0.6<\Delta\mu<1.2$, while beyond that, the junction
is operative again. The reason for this behavior is that
electronic-induced excitation and relaxation rates, $k_u^e$ and $k_d^e$,
become both small at large bias, thus the thermal bath-induced rates
dominate the mode dynamics, leading to a normal-thermal like behavior.

The coupling of the D and A molecular states to the metal leads may
be further tuned experimentally. In Fig. \ref{FigS2} we show a stability map of
the junction as a function of voltage bias and metal-molecule hybridization
$\Gamma_{L}=\Gamma_R$. We explore two situations: (a) The molecular
mode is perfectly isolated from other vibrations, and (b)
$\Gamma_{ph}$ is finite. In both cases, once metal-molecule coupling
is large enough, a stable operation sustains. This result seems
initially counterintuitive, as one expects strongly coupled
molecules to support high charge and energy currents, potentially leading to junction rupture.
 However, the
key factor in the formation of vibrational instability here is the fact
that at certain voltages the vibrational excitation rate $k_u^e$
exceeds the relaxation rate $k_d^e$. Inspecting the rates
(\ref{eq:rate2}), one can analytically prove that if the effective
density of states  is energy {\it independent},
$J_{\nu}(\epsilon)=C$, which is the case at strong metal-molecule
coupling, then $k_d^e-k_u^e \propto C^2\omega_0$, a positive number. The
key factor in instability build-up is thus the usage of electronic reservoirs
with effective DOS  [Eq. (\ref{eq:spec})]
peaked around different energies, the D and A levels.

%=====================================================

%================================

Concluding our observations: (i) We confirmed the steady-state
entropy production FT for the different model variants. This is a
non-trivial task since charge and energy currents here are not
tightly coupled, a result of the strong metal-D and A-metal
couplings. Therefore, one needs to separately count particle number
and energy transfer in the system. (ii) We derived  simple
analytical expressions for the charge current, assuming either
harmonic or an anharmonic vibrational mode. As expected,
harmonic-mode junctions better conduct since the electronic energy can
be used to excite multiple vibrational states. An anharmonic
mode quickly reaches saturation. (iii) We defined an effective
temperature for the vibration and demonstrated bias induced cooling
at low bias, $\Delta\mu<0.2$. (iv) At intermediate voltage
bias, $0.2<\Delta \mu <0.6$, heating effects were observed, and the
mode effective temperature exceeds the environmental temperature.
(v) Once the donor and acceptor levels switch position for $\Delta
\mu>0.6$, $\epsilon_d>\epsilon_a$, instability develops: The mode
excitation rate exceeds the relaxation rate, and the
vibrational mode uncontrollably heats. (vi) Coupling the vibrational
mode to an external thermal bath stabilizes the junction. In
particular, assuming a weak interaction to a phonon bath,
$\Gamma_{ph}=0.005$, junction instability is removed for
$\Delta\mu>1.2$; the electronic-induced rates diminish and the
mode dynamics is controlled by the thermal bath. (vii) The
appearance of vibrational instability can be traced down to the
metals' energy dependent DOS, different at the two ends.
%Note that the results
%(iii)-(vii) are valid for both harmonic and anharmonic modes.

The simple models described here elucidate the role of
different factors on vibrational cooling, heating, and instability build-up in a D-A
electronic rectifier.
The effects of mode frequency, its interaction with other modes, the reservoirs'
temperature, metal-molecule coupling strength,  and bias voltage,
were examined.
While the focus of this work has been on vibrational effects, 
the theory developed here
could be used for describing the coupling of an electronic junction to a cavity mode-electromagnetic 
environment. One could thus reformulate this study and describe cooling, heating and diode-like
effects in photonic heat conduction \cite{photon1,photon2,photon3,photon4,photon5}.
Future work will be devoted to the study of noise processes in phonon-assisted tunneling junctions, with the
motivation to expose mode properties (harmonicity) through the noise
characteristics.

%---------------------------------

\begin{acknowledgments}
DS acknowledges support from an NSERC discovery grant. The work of
LS  was supported by an Early Research Award of DS.
\end{acknowledgments}

%------------------------------------------------------------------------------------------

%=======================

\renewcommand{\theequation}{A\arabic{equation}}
\setcounter{equation}{0}  % reset counter
\section*{Appendix A: Full counting statistics
 for charge and energy in the dissipative anharmonic-mode rectifier model}

We describe here the derivation of the generating function and the charge current
for the anharmonic-mode bath-coupled rectifier model,
 \bea H_{A+B} &=& \frac{\omega_0}{2} \sigma_z + \sigma_x
\left(F_e+F_b\right)
\nonumber\\
&+& \sum_l  \epsilon_l a_l^{\dagger}a_l + \sum_r  \epsilon_r
a_r^{\dagger}a_r
+\sum_{\alpha}\omega_{\alpha}b_{\alpha}^{\dagger}b_{\alpha}.
\label{eq:AHBBB} \eea
Here, the TLS excitation and relaxation processes are coupled to
both electronic transitions in the junction and to a
thermal-phononic reservoir,
\bea F_e&=&\kappa \sum_{l,r} (\lambda_l^* \lambda_r a_l^{\dagger}a_r
+ \lambda_r^* \lambda_l a_r^{\dagger}a_l). \nonumber\\
F_b&=&\sum_{\alpha} v_{\alpha}(b_{\alpha}^{\dagger}+b_{\alpha}).
\eea
$a_l$, $a_r$ are fermionic annihilation operators corresponding to
the left and right metals; the coefficients $\lambda_{\nu}$ were
defined in Eq. (\ref{eq:diag}),  $\kappa$ denotes the D-A
tunneling strength, $b_{\alpha}$ is a bosonic operator, describing
the $\alpha$ reservoir mode, $v_{\alpha}$ quantifies the TLS-bath
interaction strength. For more details, see text around Eq.
(\ref{eq:HBBB}).

The impurity (TLS) dynamics can be obtained by using a
master-equation approach \cite{Breuer}. The procedure involves a
second order perturbation theory in the impurity coupling to both
electronic and phononic reservoirs. In the markovian limit, under the rotating wave approximation, one
standardly achieves kinetic equations that separately
account for the electronic (e) and phononic (b) relaxation pathways,
\bea
&&\dot p_1=-\left( k_{1\rightarrow 0}^e + k_{1\rightarrow 0}^b \right)p_1
+ \left(k_{0\rightarrow 1}^e +  k_{0\rightarrow 1}^b\right)p_0
\nonumber\\
&&p_1+p_0=1.
\label{eq:ADTLS}
\eea
The relaxation terms are given by
\bea
k_{n\rightarrow n'}^e &=& \int_{-\infty}^{\infty}
e^{i(\epsilon_n-\epsilon_{n'}) \tau}
\langle F_e(\tau) F_e(0)\rangle d\tau
\nonumber\\
k_{n\rightarrow n'}^b&=& \int_{-\infty}^{\infty}
e^{i(\epsilon_n-\epsilon_{n'}) \tau} \langle F_b(\tau) F_b(0)\rangle
d\tau. \label{eq:AkTLS} \eea
Here, $\epsilon_n$ is the energy of the $n$th vibrational level.
Electron induced rate constants are detailed through Eqs.
(\ref{eq:kTLS})-(\ref{eq:rate2}). The thermal bath induced rates can
be similarly put together,
\bea
k_{1\rightarrow0}^b&=& \Gamma_{ph}(\omega_0)[n_{ph}(\omega_0)+1],
\nonumber\\
k_{0\rightarrow1}^b&=&
 k_{1\rightarrow0}^b e^{-\omega_0\beta_{ph}}
%\Gamma_{ph}(\omega_0)n_{ph}(\omega_0)
\eea
$n_{ph}(\omega)=[e^{\beta_{ph}\omega}-1]^{-1}$ denotes the
Bose-Einstein distribution function and
$\Gamma_{ph}(\omega)=2\pi\sum_{\alpha}v_{\alpha}^2
\delta(\omega_{\alpha}-\omega)$. For brevity, we ignore below the
direct reference to frequency. We now define the probability
distribution function $\mathcal P_t(n,N,\omega_L,\omega_R,q\omega_0)$, as the
probability to find the system at time $t$ in state $n=0,1$, with
$N$ electrons transferred to the right bath, $\omega_{\nu}$ excess
energy accumulated at the $\nu$ bath ($\nu=L,R$), and $q\omega_0$
energy attained by the phonon bath, due to the transfer of $q$
quantas from the TLS to this bath. Note that charge conservation between the $L$
and $R$ baths is enforced, allowing us to work with a single counting
field for describing charge transfer processes in the steady-state
limit.

We now resolve the associated master equation for the two-state
population, to its charge and energy contributions. The resulting equations
are analogous to Eq. (\ref{eq:eqP}),
\begin{widetext}
\bea
 \dot{\mathcal  P}_t(1,N,\omega_L,\omega_R,q\omega_0)&=&
- \mathcal P_t(1,N,\omega_L,\omega_R,q\omega_0) \left[k_{1\rightarrow 0}^e +
\Gamma_{ph}(\omega_0)[n_{ph}(\omega_0)+1]\right]
\nonumber\\
&+& \intinf \mathcal P_t(0,N-1,\omega_L+ \epsilon, \omega_R-\epsilon+\omega_0,q\omega_0)
f_L(\epsilon) [1-f_R(\epsilon-\omega_0)]
J_L(\epsilon)J_R(\epsilon-\omega_0) d\epsilon
\nonumber\\
&+&\intinf \mathcal P_t(0,N+1, \omega_L-\epsilon+\omega_0, \omega_R+\epsilon,q\omega_0) f_R(\epsilon)
[1-f_L(\epsilon-\omega_0)] J_R(\epsilon)J_L(\epsilon-\omega_0)
d\epsilon \nonumber\\
&+& \mathcal P_t(0,N, \omega_L,\omega_R,(q+1)\omega_0) \Gamma_{ph}(\omega_0) n_{ph}(\omega_0)
\nonumber\\
 \dot{\mathcal  P}_t(0,N,\omega_L,\omega_R,q\omega_0)&=&
- \mathcal P_t(0,N,\omega_L,\omega_R,q\omega_0) \left[k_{0\rightarrow 1}^e + \Gamma_{ph}(\omega_0)n_{ph}(\omega_0) \right]
\nonumber\\
&+& \intinf \mathcal P_t(1,N-1, \omega_L+\epsilon, \omega_R -\epsilon-\omega_0,q\omega_0)
f_L(\epsilon) [1-f_R(\epsilon+\omega_0)]
J_L(\epsilon)J_R(\epsilon+\omega_0) d\epsilon
\nonumber\\
&+&\intinf \mathcal P_t(1,N+1, \omega_L-\epsilon-\omega_0,\omega_R+\epsilon,q\omega_0) f_R(\epsilon)
[1-f_L(\epsilon+\omega_0)] J_R(\epsilon)J_L(\epsilon+\omega_0)
d\epsilon \label{eq:AeqP}
\nonumber\\
&+& \mathcal P_t(1,N, \omega_L,\omega_R,(q-1)\omega_0) \Gamma_{ph}(\omega_0) [n_{ph}(\omega_0)+1].
\eea
\end{widetext}
We Fourier transform this system with respect to charge and
energy,
\bea
&& \mathcal P_t(n,\chi,\eta_L,\eta_R,\xi) =
\nonumber\\
&&
\sum_{N=-\infty}^{\infty} e^{iN\chi}
\sum_{q=-\infty}^{\infty} e^{iq\omega_0\xi}
\intinf  e^{i\omega_L\eta_L}\,d\omega_L
\nonumber\\
&&\times
\intinf  e^{i\omega_R\eta_R}\,d\omega_R
\mathcal P_t(n,N,\omega_L,\omega_R,q\omega_0)
\eea
to obtain the characteristic function
$\mathcal{Z}(\chi,\eta_L,\eta_R,\xi,t)$. It depends on the energy
counting fields $\eta_{\nu}$ and $\xi$, and the charge counting field
$\chi$. It is a vector with two entries, as in Eq. (\ref{eq:z}), and
it satisfies the differential equation
\beq \d{\ket{\mathcal Z}}{t} = - \mathcal{\hat W}\ket{\mathcal Z}.
\label{eq:AZ} \eeq
The matrix $\mathcal{\hat W}$ has the following entries
\begin{widetext}
\bea &&\mathcal{\hat W} =
\nonumber\\
&&\begin{pmatrix}
k_{0\rightarrow 1}^e +\Gamma_{ph}n_{ph}  & -e^{i\chi}F_1^-(\eta_{L},\eta_R)-e^{-i\chi}F_2^+(\eta_L,\eta_R)
 -\Gamma_{ph} (n_{ph}+1)e^{i\xi\omega_0} \\
-e^{i\chi}F_1^+(\eta_L,\eta_R)-e^{-i\chi}F_2^-(\eta_L,\eta_R) - \Gamma_{ph}n_{ph}e^{-i\xi\omega_0}  & k_{1\rightarrow 0}^e +\Gamma_{ph}(n_{ph}+1)  \\
\end{pmatrix}
\nonumber\\ \label{eq:Amu} \eea
\end{widetext}
where
\bea F_1^{\pm}(\eta_L,\eta_R)&=&
\intinf e^{-i\epsilon\eta_L} e^{i(\epsilon\mp\omega_0)\eta_R}
\nonumber\\
&\times&
f_L(\epsilon)[1-f_R(\epsilon\mp\omega_0)]J_L(\epsilon)J_R(\epsilon\mp\omega_0)d\epsilon
\nonumber\\
\eea
and
\bea
F_2^{\pm}(\eta_L,\eta_R)&=& \intinf
e^{i\eta_L(\epsilon\pm\omega_0)}e^{-i\epsilon\eta_R}
\nonumber\\
&\times&[1-f_L(\epsilon\pm\omega_0)]f_R(\epsilon)J_L(\epsilon\pm\omega_0)J_R(\epsilon)d\epsilon.
\nonumber\\
\eea
The CGF is expressed in terms of the characteristic function $|\mathcal Z\rangle$ as
\bea
G(\chi,\eta_L,\eta_R,\xi) =  \lim_{t \to \infty} \ \frac{1}{t}\ln \langle I| \mathcal Z \rangle,
\eea
Practically, it is reached by the negative of the smallest
eigenvalue of the matrix $\mathcal{\hat W}$,
\bea
G(\chi,\eta_L,\eta_R,\xi) &=& -\frac{w_{1,1}+w_{2,1}}{2} 
\\
&+& \frac{\sqrt{(w_{1,1}-w_{2,2})^2 +4w_{1,2}w_{2,1}}}{2},
\nonumber
\eea
with $w_{i,j}$ the matrix elements in Eq. (\ref{eq:Amu}).
The charge current is obtained by taking the first derivative with respect to $(i\chi)$,
\bea
\avg{I_e}&=&\frac{\partial G}{\partial i\chi}\Bigg|_{\chi,\eta_{\nu},\xi=0}
= (w_{1,1}+w_{2,2})^{-1} 
\nonumber\\
&\times&
\left[  w_{2,1}\frac{\partial w_{1,2}  } {\partial i\chi} 
        +w_{1,2}\frac{\partial w_{2,1}  } {\partial i\chi} \right]
\Bigg|_{\chi,\eta_{\nu,\xi=0}},
\eea
where the following holds when the counting fields are all set to zero,
$w_{1,2}=-w_{2,2}$, $w_{2,1}=-w_{1,1}$, $\partial w_{1,2}/\partial(i\chi)|_{0}= k_{1\rightarrow0}^{R\rightarrow L}
- k_{1\rightarrow0}^{L\rightarrow R}$ and
$\partial w_{2,1}/\partial(i\chi)|_{0}= k_{0\rightarrow1}^{R\rightarrow L}- k_{0\rightarrow1}^{L\rightarrow R}$.
Recall that the rates $k_{n\rightarrow n'}^{\nu\rightarrow \nu'}$ are electron-hole generation assisted rates,
see Eq. (\ref{eq:rate2}).
We can now identify the levels population,
\bea  \frac{w_{1,1}}{w_{1,1}+w_{2,2}} &=&
\frac{ k_{0\rightarrow1}^e + \Gamma_{ph} n_{ph}} {
k_{0\rightarrow1}^e+k_{1\rightarrow0}^e+ \Gamma_{ph}[2n_{ph}+1]} 
\nonumber\\
&=&p_1,
\eea
and similarly  for $p_0=1-p_1$, resulting in the expression for the charge current
\bea
\avg{I_e}=p_1(k_{1\rightarrow0}^{L\rightarrow R}- k_{1\rightarrow0}^{R\rightarrow L})
+p_0 (k_{0\rightarrow1}^{L\rightarrow R}- k_{0\rightarrow1}^{R\rightarrow L}).
\eea
This result is formally identical to Eq. (\ref{eq:Ie}),
with the only difference that the steady-state TLS population is now modified, to
include phonon-bath assisted transitions.
%
%========================

\end{document}